\documentclass[a4paper,11pt]{article}
\usepackage{heppub}

\usepackage{xspace}
\usepackage{placeins}
\usepackage{wasysym}
\usepackage{slashed}
\usepackage{empheq}
\usepackage[font={bf, footnotesize}]{subfig}
\usepackage[usenames,dvipsnames]{xcolor}

\newcommand{\abs}[1]{\lvert#1\rvert}
\newcommand{\Abs}[1]{\bigl\lvert#1\bigr\rvert}

\newcommand{\ord}[1]{\mathcal{O}(#1)}
\newcommand{\Ord}[1]{\mathcal{O}\bigl(#1\bigr)}
\newcommand{\ORd}[1]{\mathcal{O}\Bigl(#1\Bigr)}
\newcommand{\ordsq}[1]{\mathcal{O}[#1]}
\newcommand{\Ordsq}[1]{\mathcal{O}\bigl[#1\bigr]}

\newcommand{\Mae}[3]{\big\langle#1\big\lvert#2\big\rvert#3\big\rangle}

\newcommand{\ket}[1]{\lvert#1\rangle}
\newcommand{\Ket}[1]{\big\lvert#1\big\rangle}

\newcommand{\Bra}[1]{\big\langle#1\big\rvert}

\newcommand{\nn}{\nonumber}

\newcommand{\df}{\mathrm{d}}
\newcommand{\img}{\mathrm{i}}

\renewcommand{\Re}{\operatorname{Re}}

\newcommand{\tr}{\operatorname{tr}}
\newcommand{\Tr}{\operatorname{Tr}}

\newcommand{\Sl}[1]{\slashed{#1}}

\newcommand{\ktcut}{{k_T^\mathrm{cut}}}
\newcommand{\xcut}{x_\mathrm{cut}}

\DeclareMathOperator*{\SumInt}{%
\mathchoice%
  {\ooalign{$\displaystyle\sum$\cr\hidewidth$\displaystyle\int$\hidewidth\cr}}
  {\ooalign{\raisebox{.14\height}{\scalebox{.7}{$\textstyle\sum$}}\cr\hidewidth$\textstyle\int$\hidewidth\cr}}
  {\ooalign{\raisebox{.2\height}{\scalebox{.6}{$\scriptstyle\sum$}}\cr$\scriptstyle\int$\cr}}
  {\ooalign{\raisebox{.2\height}{\scalebox{.6}{$\scriptstyle\sum$}}\cr$\scriptstyle\int$\cr}}
}

\newcommand{\eps}{\epsilon}

\newcommand{\w}{\omega}

\newcommand{\cA}{\mathcal{A}}

\newcommand{\cH}{\mathcal{H}}

\newcommand{\cL}{\mathcal{L}}
\newcommand{\cM}{\mathcal{M}}

\newcommand{\cP}{\mathcal{P}}
\newcommand{\cR}{\mathcal{R}}

\newcommand{\cJ}{\mathcal{J}}
\newcommand{\cS}{\mathcal{S}}

\newcommand{\fh}{\mathfrak{h}}

\newcommand{\tcR}{\tilde{\cR}}

\newcommand{\bn}{{\bar n}}

\newcommand{\as}{\alpha_s}

\newcommand{\lqcd}{\Lambda_\mathrm{QCD}}
\newcommand{\MSbar}{\overline{\text{MS}}\xspace}
\newcommand{\nbar}{\bar{n}}
\newcommand{\zb}{\bar{z}}

\newcommand{\cut}{\mathrm{cut}}

\newcommand{\bare}{\mathrm{bare}}
\newcommand{\UV}{\mathrm{UV}}

\newcommand{\WidthThreeSubfigs}{0.32\textwidth}

\allowdisplaybreaks[2]

\title{\boldmath Transverse Momentum-Dependent Heavy-Quark Fragmentation at Next-to-Leading Order}

\author[a]{Rebecca von Kuk,}
\emailAdd{rebecca.von.kuk@desy.de}

\author[b,c,d]{Johannes K.~L.~Michel,}
\emailAdd{j.k.l.michel@uva.nl}

\author[b]{and Zhiquan Sun}
\emailAdd{zqsun@mit.edu}

\affiliation[a]{Deutsches Elektronen-Synchrotron DESY, Notkestr. 85, 22607 Hamburg, Germany}
\affiliation[b]{Center for Theoretical Physics,\,Massachusetts Institute of Technology,\,Cambridge,\,MA\,02139,\,USA}
\affiliation[c]{Institute for Theoretical Physics Amsterdam and Delta Institute for Theoretical Physics, University of Amsterdam, Science Park 904, 1098 XH Amsterdam, The Netherlands}
\affiliation[d]{Nikhef, Theory Group, Science Park 105, 1098 XG, Amsterdam, The Netherlands}

\abstract{%
The transverse momentum-dependent fragmentation functions (TMD FFs)
of heavy (bottom and charm) quarks, which we recently introduced,
are universal building blocks that enter predictions for a large number of observables
involving final-state heavy quarks or hadrons.
They enable the extension of fixed-order subtraction schemes to quasi-collinear limits,
and are of particular interest in their own right
as probes of the nonperturbative dynamics of hadronization.
In this paper we calculate all TMD FFs involving heavy quarks
and the associated TMD matrix element in heavy-quark effective theory (HQET)
to next-to-leading order in the strong interaction.
Our results confirm the renormalization properties,
large-mass, and small-mass consistency relations predicted in our earlier work.
We also derive and confirm a prediction for the large-$z$ behavior
of the heavy-quark TMD FF by extending, for the first time,
the formalism of joint resummation
to capture quark mass effects in heavy-quark fragmentation.
Our final results in position space agree with those of a recent calculation
by another group that used a highly orthogonal organization
of singularities in the intermediate momentum-space steps,
providing a strong independent cross check.
As an immediate application, we present the complete quark mass dependence
of the energy-energy correlator (EEC) in the back-to-back limit
at $\mathcal{O}(\alpha_s)$.
}

\date{April 25, 2024}

\preprint{\vbox{%
\hbox{MIT-CTP 5643}
\hbox{DESY-23-174}
\hbox{Nikhef 2023-018}
}
}

\begin{document}

\maketitle
\newpage

\section{Introduction}
\label{sec:intro}

The fragmentation of heavy (bottom or charm) quarks into
the experimentally observed heavy meson and baryon states
is of particular interest
because the mass of the quark imprints as a perturbative scale
on the otherwise nonperturbative process of hadronization.
In \refcite{vonKuk:2023jfd}, we initiated the study
of transverse momentum-dependent (TMD) fragmentation functions (FFs)
for the formation of heavy hadrons from a parent heavy quark.
Our work generalized the well-studied case
where only the collinear momentum fraction of the hadron is resolved~\cite{Mele:1990cw, Jaffe:1993ie,Falk:1993rf,Neubert:2007je, Fickinger:2016rfd}
(see \refscite{Bonino:2023icn} for a recent precision study on data),
and is part of a larger, ongoing research effort to understand
differential jet and fragmentation functions
involving heavy quarks, heavy hadrons,
and heavy-quark bound states~\cite{Makris:2018npl, Echevarria:2020qjk, vonKuk:2023jfd,
Copeland:2023wbu, Echevarria:2023dme, Copeland:2023qed, Dai:2023rvd, Caletti:2023spr}.
In this context, TMD FFs are unique in the wealth of information
they can provide on the hadronization process,
essentially offering a full three-dimensional view of the fragmentation cascade.
The main goal of the present paper is to evaluate all unpolarized heavy-quark TMD FFs
at perturbative transverse momenta to complete next-to-leading order (NLO)
in the strong coupling,
with the aim of improving the baseline perturbative precision
for studying
their rich nonperturbative structure.

Importantly, the factorization paradigm ensures
that the heavy-quark TMD FFs appear as universal building blocks
across predictions for a large number of processes
involving final-state heavy quarks.
In \refcite{vonKuk:2023jfd}, we explicitly considered their phenomenology
in heavy-hadron pair production in the back-to-back limit in $e^+e^-$ collisions
and in (polarized) semi-inclusive DIS at the future Electron-Ion Collider~\cite{AbdulKhalek:2021gbh},
cf.\ \refcite{Hekhorn:2024tqm} for recent dedicated projections
for polarized \emph{collinear} parton distribution functions (PDFs)
and FFs involving heavy quarks at the EIC.
However, the applicability of heavy-quark TMD FFs is much wider:
Applications of immediate phenomenological interest
range from $Z$$+$hadron~\cite{Chien:2022wiq} and dihadron~\cite{Gao:2023ulg}
azimuthal decorrelations in $pp$ and $pA$ collisions
to hadron-in-jet transverse-momentum distributions~\cite{Kang:2017glf, Kang:2020xyq},
transverse momentum-like event shapes
(extending the calculation of mass effects in thrust-like event shapes~\cite{Hoang:2019fze, Lepenik:2019jjk, Bris:2020uyb, Bris:2022cdr}),
and energy-energy correlators in the back-to-back limit
in $e^+e^-$~\cite{Moult:2018jzp}, $ep$ and $eA$~\cite{Li:2021txc, Kang:2023big},
and $pp$ collisions~\cite{Gao:2019ojf, Gao:2023ivm}.
In all of these cases, factorization formulas involving TMD FFs
have been derived, and the NLO results we present in this paper
enable one to fully account for the effect of quark masses
and accurately capture their highly nontrivial
interplay with transverse momenta in all of these processes.

\begin{figure*}
\centering
\includegraphics[width=0.8\textwidth]{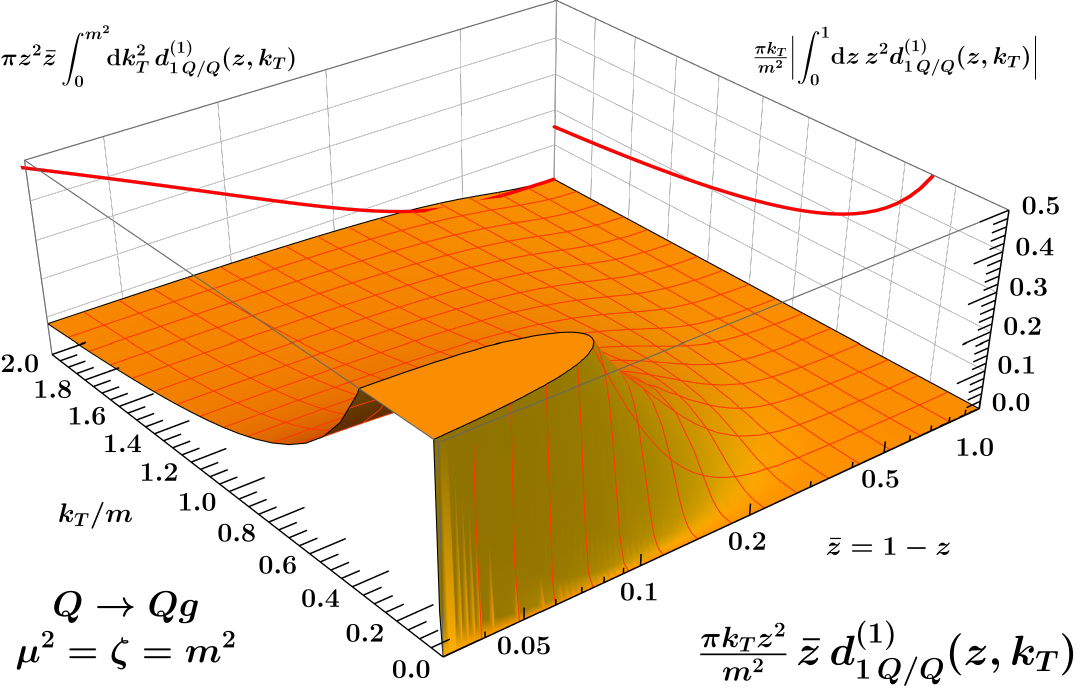}
\caption{
The probability density $\df \cP_{Q\bar{Q} \to (Qg)\bar{Q}}$
for a gluon to split off a heavy quark in the quasi-collinear limit,
which coincides with the heavy-quark TMD FF $d_{1\,Q/Q}$
up to a factor of $z^2$
for finite collinear gluon momentum fraction $\zb > 0$
and transverse momentum $k_T > 0$.
On the rear walls of the plot we show the integrals of $d_{1\,Q/Q}$
over one of $k_T$ or $z$ as predicted by the NLO calculation we perform in this paper,
showcasing that the TMD regularization and renormalization procedure
renders them finite
despite the intricate singular behavior of the integrand.
The same is true for the total integral over both $z$ and $k_T < m$,
which is not shown in the plot,
but at the values of the scales chosen here
is equal to $\frac{\as C_F}{4\pi} \bigl( 5 - \frac{\pi}{2} - \frac{2\pi^2}{3} \bigr) \approx -0.067$.
For numerical illustration, we take $\as(\mu = m) = 0.2$.
}
\label{fig:3d_plot_d1qq}
\end{figure*}

The heavy-quark TMD FF is intimately connected
with the universal gluon splitting probability off a heavy quark
in the quasi-collinear limit~\cite{Catani:2000ef, Cacciari:2002xb},
e.g.\ after heavy-quark pair production has taken place,
\begin{align} \label{eq:heavy_quark_splitting_probability}
\frac{\df \cP_{Q\bar{Q} \to (Qg)\bar{Q}}}{\df z \, \df (k_T^2) }
= \frac{\as C_F}{2\pi} \, \frac{
   k_T^2 z^2 (1 + z^2) + m^2(1-z)^4
}{
   (1-z)\bigl[ k_T^2 z^2 + m^2 (1-z)^2\bigr]^2
}
\,,\end{align}
where $m$ is the heavy-quark mass,
$k_T$ is the transverse momentum of the gluon
with respect to the heavy quark,
and $z$ is the collinear momentum fraction retained by the heavy quark.
The quasi-collinear splitting amplitude is shown
in a three-dimensional plot as a function of $k_T$
and the momentum fraction $\zb \equiv 1 - z$ carried by the gluon
in \fig{3d_plot_d1qq}.
Famously, and in contrast to the behavior of light partons,
the splitting amplitude is suppressed
for energetic gluons with large momentum fractions $\zb \gtrsim 0.2$,
an effect known as the dead cone~\cite{Dokshitzer:1991fd}.
The splitting probability is strongly peaked at small $\zb$ and small $k_T \lesssim m$.
Unlike for light partons,
it eventually turns over for any finite value of $\zb > 0$
and vanishes (near the front right edge of the plot) as $k_T \to 0$,
yet the simultaneous limit $\zb, k_T \to 0$ diverges.
A key challenge we address in this paper is how to consistently extend
this complicated singular behavior of the quasi-collinear splitting probability
into the limits $k_T \to 0, \zb \to 0$ where the gluon is unresolved.
The right framework for doing so is -- indeed -- TMD factorization~\cite{Collins:1350496},
which provides precise field-theoretic prescriptions
to isolate, regulate, and renormalize the associated divergences,
and it is precisely in this sense that the heavy-quark TMD FF
is the formal generalization of \eq{heavy_quark_splitting_probability}
to $\zb \to 0$ and/or $k_T \to 0$
in a way that is manifestly finite in four spacetime dimensions.
As a demonstration, we include on the rear walls
of the three-dimensional plot in \fig{3d_plot_d1qq}
the integral of the heavy-quark TMD FF over one of $z$ or $k_T \leq m$,
respectively, as predicted by our calculations in this paper,
showcasing that they are indeed individually finite.
Similarly, the total integral over both variables $0 \leq z \leq 1$
and $0 \leq k_T \leq m$ (and thus any integral
over a subset of the plane) is also finite in the usual distributional sense,
and for the indicated values of the scales simply evaluates to $\frac{\as C_F}{4\pi} \bigl( 5 - \frac{\pi}{2} - \frac{2\pi^2}{3} \bigr)$.
We stress that isolating this singular behavior at fixed order
in terms of a universal and finite heavy-quark TMD FF
is not an academic exercise,
but instead is of immediate interest to extend fixed-order subtraction
schemes like that of $q_T$ subtractions~\cite{Catani:2007vq}
and its multi-jet generalizations~\cite{Gaunt:2015pea, Bertolini:2017efs, Buonocore:2022mle, Buonocore:2023rdw}
to quasi-collinear limits near final-state heavy quarks.
Such extensions, which can potentially yield large gains in numerical efficiency,
are likewise an active topic of current research, see e.g.\ \cite{Dhani:2023uxu, Craft:2023aew, AH:2024ueu, Zidi:2024lid}.

The remainder of this paper is structured as follows:
\Sec{theoretical_framework} contains a review of the theoretical framework introduced in \refcite{vonKuk:2023jfd}
to describe heavy-quark fragmentation in the transverse plane.
In \sec{partonic_heavy_quark_tmd_ff},
we calculate all heavy-quark TMD FFs
to next-to-leading order in the strong coupling,
and present the results in both momentum and position space.
In \sec{bhqet_tmd_fragmentation_factors}, we perform an analogous calculation for the effective TMD matrix elements
that arise in boosted heavy-quark effective theory (bHQET)
in the limit where the transverse momentum is small compared to the mass.
We employ these results and known one-loop results in the literature
to verify the expected small and large-mass consistency relations in \sec{consistency_relations}.
\Sec{joint_res} is dedicated to an extension of the joint resummation formalism
to massive quark fragmentation, which explains
the surprising simplicity of our results at large $z$.
As an immediate, first application of our results,
we present the complete mass dependence
of the energy-energy correlator in the back-to-back limit in \sec{eec}.
In \sec{comparison_to_2310_19207}, we compare our main results to those of \refcite{Dai:2023rvd},
where an independent calculation of the one-loop heavy-quark TMD FF
was recently performed.
We find full agreement at the level of the final position-space results,
but stress the differences in how the intermediate momentum-space calculations
are organized (amounting to a strong independent cross check)
and -- on the phenomenological side --
the differences in how the $z$ dependence is treated in the bHQET limit.
We conclude in \sec{conclusions}.

\section{Theoretical framework}
\label{sec:theoretical_framework}

\subsection{Heavy-quark TMD FFs}
\label{sec:theory_intro}

We study the fragmentation of a heavy quark $Q$ with pole mass $m \gg \lqcd$
into an unpolarized heavy hadron $H$ and additional fragmentation products $X$.
We work in QCD with $n_f=n_\ell+1$ flavors,
where $n_\ell$ is the number of massless quark flavors.
We work in the ``hadron frame'' for fragmentation~\cite{Collins:1350496}
where $P_H^\mu = (P_H^-, M_H^2/P_H^-, 0)$,
with $M_H$ the hadron mass,
and $P_{H,\perp}=0$ by definition.
(We refer to \App{lc_coordinates} for our lightcone conventions.)

In \refcite{vonKuk:2023jfd} we showed
that up to a factor of the total probability $\chi_H$ for $Q$ to fragment into $H$,
the distribution differential in $k_T = P_{X,T} \sim M_H$
and the lightcone momentum fraction $z_H$ of the original quark retained by $H$
is governed by a new perturbative matching coefficient,
the \emph{partonic heavy-quark TMD FF} $d_{1\,Q/Q}(z, k_T, \mu, \zeta)$,
\begin{align} \label{eq:tmd_ff_unpol_bhqet}
D_{1\,H/Q}(z_H, b_T, \mu, \zeta)
= d_{1\,Q/Q}(z_H, b_T, \mu, \zeta) \, \chi_H
+ \ORd{\frac{\lqcd}{m}}
+ \ord{\lqcd b_T}
\,.\end{align}
Making all regulators explicit,
the formal definition of $d_{1\,Q/Q}$
at the bare level reads
\begin{align} \label{eq:def_d1qq_bT_bare}
d_{1\,Q/Q}^\bare(z, b_T, \eps, \eta, \zeta/\nu^2)
&=
\frac{1}{2 z^{1-2\eps} N_c} \int \! \frac{\df b^+}{4\pi} \, e^{\img b^+ (p^-/z)/2}
\\[0.4em] & \quad \times
\Tr \SumInt_{X}
\tr \Bigl[ \frac{\Sl{\bn}}{2} \,
\Mae{0}{W_\eta^\dagger(b) \, \psi_Q(b)}{Q X}
\Mae{Q X}{\bar\psi_Q (0) \, W_\eta(0)}{0}
\Bigr]
\nn \,,\end{align}
where we work in $d = 4-2\eps$ spacetime dimensions,
$N_c$ is the number of colors,
$\Tr$ ($\tr$) indicates a trace over color (spin),
and $\SumInt_{X}$ indicates a sum over all possible partonic final states
combined with an integral over their phase space.
The fields in the first matrix element are evaluated
at a spacetime position $b \equiv (0, b^+, b_\perp)$
with $b_\perp$ Fourier conjugate to $k_\perp$,
and $b_T^2 \equiv -b_\perp^2$ and $k_T^2 = -k_\perp^2$, respectively.
The heavy quark in the external state carries momentum
$p^\mu = (p^-, m^2/p^-, 0)$,
i.e., the above definition is equal to the hadron-level definition
of the heavy-quark TMD FF $D_{1\,H/Q}$,
but with $H$ replaced by $Q$ itself in the external state
and restricting to partonic final states $X$~\cite{vonKuk:2023jfd}.
The definition in \eq{def_d1qq_bT_bare} is given
in terms of the heavy-quark field renormalized on shell,
\begin{align} \label{eq:quark_field_ren}
\psi^\bare_Q(x) = Z_{\psi,\mathrm{OS}}^{1/2}(m, \mu, \eps) \, \psi_Q(x)
\,,\end{align}
and in terms of Wilson lines $W_\eta(x)$ defined as anti-path ordered exponentials of gauge fields
extending to positive infinity along the lightcone direction $\nbar^\mu$,
\begin{align} \label{eq:def_wilson_line}
W_\eta(x) = \bar{P} \Bigl[ \exp \Bigl(
   - \img g \int_0^\infty \! \df s \, \bn \cdot A(x + \bn s)
\Bigr) \Bigr]_\eta
\,.\end{align}
The subscript $\eta$ indicates the presence of an additional rapidity regulator.
In explicit calculations,
we will use the so-called $\eta$ regulator of \refscite{Chiu:2011qc, Chiu:2012ir}
whose action on a single gluon with momentum $\ell$ attached to the Wilson line
amounts to inserting a factor of $\abs{\ell^-/\nu}^\eta$
under phase-space or loop integrals over $\ell$,
with $\nu$ a dimensionful rapidity scale.
The final result for the bare matrix element depends on $\zeta/\nu^2$,
with $\zeta \equiv (p^-/z)^2$ the Collins-Soper scale.

In terms of the bare collinear matrix element in \eq{def_d1qq_bT_bare},
the renormalized partonic heavy-quark TMD FF is given by
\begin{align} \label{eq:def_d1qq_bT_ren}
d_{1\,Q/Q}(z, b_T, \mu, \zeta)
&= \lim_{\eps \to 0} Z_\UV(\mu, \zeta, \eps)
\lim_{\eta \to 0} \Bigl[
   d_{1\,Q/Q}^\bare(z, b_T, \eps, \eta, \zeta/\nu^2) \, \sqrt{S}(b_T, m, \eps, \eta, \nu)
\Bigr]
\,,\end{align}
where $S$ is the universal bare TMD soft function for the $\eta$ regulator,
which cancels all poles of $\eta$ and the associated $\nu$ dependence,
and is independent of the heavy-quark mass
up to secondary quark mass effects starting at two loops.
By contrast, the $\MSbar$ renormalization factor $Z_\UV$
for quark TMDs is independent of the mass to all orders
by RG consistency with the hard matching coefficient
at the hard scattering energy $Q \sim p^-/z \gg m \sim k_T \sim 1/b_T$.
While soft subtractions (known as zero bins in the SCET literature)
generally need to be accounted for in both virtual and real collinear diagrams
when computing the bare collinear matrix element itself,
we have explicitly verified that
they lead to scaleless integrals for our choice of regulator,
also in the presence of the quark mass.

\subsection{Nonvalence contributions}
\label{sec:theory_nonvalence}

There are two further ways in which heavy quarks can participate
in the TMD fragmentation process for $m \sim k_T \sim 1/b_T \gg \lqcd$;
these were sketched in \refcite{vonKuk:2023jfd}, but we spell them out here explicitly.
In one case, the resolved final state is a heavy hadron $H$,
while the parent parton $i = g, q, \bar{q}$ is light,
\begin{align} \label{eq:nonvalence_matching_heavy_hadron}
D_{1\,H/i}(z_H, b_T, \mu, \zeta)
= d_{1\,Q/i}(z_H, b_T, \mu, \zeta) \, \chi_H
+ \ord{\lqcd}
\,.\end{align}
Here $\chi_H$ is again the universal inclusive fragmentation probability for $Q \to H$,
while the bare matching coefficient $d_{1\,Q/i}$ is obtained
from \eqs{def_d1qq_bT_bare}{def_d1qq_bT_ren}
after replacing the heavy-quark fields in the bare correlator
by suitable unpolarized combinations of gluon or light-quark fields.

In the other case, the resolved final-state hadron $h$ is light.
In this case the TMD FF $D_{1\,h/i}$ for $i = g, q, \bar{Q}, Q, \bar{Q}$
has to be matched onto collinear fragmentation functions $D_{h/j}$
at the scale $\lqcd$ where all degrees of freedom $j = g, q, \bar{q}$ are light,
\begin{align} \label{eq:nonvalence_matching_light_hadron}
D_{1\,h/i}(z_H, b_T, \mu, \zeta)
= \frac{1}{z_H^2} \sum_j \int \! \frac{\df z}{z} \,
\cJ_{j/i}(z, b_T, m, \mu, \zeta) \,
D_{h/j}\Bigl( \frac{z_H}{z}, \mu \Bigr)
+ \ord{\lqcd}
\,.\end{align}
This has the standard form of matching TMD FFs onto collinear FFs,
but note that we include the mass as an additional third argument of the matching coefficient
to make explicit that even when $i$ is light, the heavy quark
may in general contribute to the matching at $\mu \sim m \sim k_T$
through closed loops.

\subsection{bHQET fragmentation factors and the large-mass limit}
\label{sec:theory_large_mass}

In \refcite{vonKuk:2023jfd}, we showed that for transverse momenta $k_T \ll m$,
the TMD dynamics are governed by new (and in general nonperturbative)
matrix elements defined in boosted HQET~\cite{Fleming:2007qr, Fleming:2007xt}
that we dubbed TMD fragmentation factors.
Specifically, for $\lqcd \lesssim k_T \ll m$ and counting
\begin{align} \label{eq:how_we_count_zH}
1 - z_H \sim 1
\,,\end{align}
which amounts to integrating over wide bins in $z_H$
or taking low $z_H$ moments,
the TMD FF for producing a heavy hadron $H$ off a heavy quark $Q$
factorizes as~\cite{vonKuk:2023jfd}
\begin{align} \label{eq:tmdffs_bhqet_all_order_final_results}
D_{1\,H/Q}(z_H, b_T, \mu, \zeta)
= \delta(1-z_H) \, C_m(m, \mu, {\zeta}) \, \chi_{1,H}\Bigl(b_T, \mu, {\frac{\sqrt{\zeta}}{m}}\Bigr) + \ORd{\frac{1}{m}}
\,,\end{align}
where $C_m(m, \mu, \zeta)$ is generated by separately
matching collinear and soft modes at the scale $\mu \sim m$ onto HQET
and QCD with $n_\ell$ flavors, respectively~\cite{Hoang:2015vua};
explicit expressions for $C_m$ and other perturbative ingredients
to the order required for our perturbative checks
are given in \app{perturbative_ingredients}.
We note that \eq{how_we_count_zH} may appear counterintuitive,
given that the prediction of HQET in \eq{tmdffs_bhqet_all_order_final_results}
-- after assuming this counting -- comes out proportional to $\delta(1-z_H)$.
The proper way to understand \eq{how_we_count_zH} in this context
is that it may be satisfied not just point by point in $z_H$
far from the endpoint (where HQET predicts that the heavy-quark TMD FF is suppressed),
but in particular is also satisfied
when integrating over a wide bin in $z_H$ from $1$ down to some bin boundary $z_H^\cut$
with $1 - z_H^\cut \sim 1$,
or when taking Mellin moments $z_H^N$ with $N \ll m/\lqcd$ not large.
In these latter two cases, the prediction from HQET proportional to $\delta(1 - z_H)$
can be easily understood as all bin integrals and Mellin moments
becoming equal and independent of $z_H^\cut$ and $N$ up to corrections of $\ord{1/m}$~\cite{vonKuk:2023jfd}.
We point out that the assumption in \eq{how_we_count_zH}
in fact also underlies the ``full theory'' factorization for $k_T \sim m$
in \eq{tmd_ff_unpol_bhqet}.
Lowering $1 - z_H \ll 1$ at fixed $k_T \sim m$ in that case
requires further resummation and -- in the extreme limit --
a different factorization of nonperturbative physics,
which we present in \Sec{joint_res}.

In \eq{tmdffs_bhqet_all_order_final_results}, the TMD dynamics are encoded
in the unpolarized TMD fragmentation factor $\chi_{1,H}$.
For $\lqcd \ll k_T \sim 1/b_T$, we showed in \refcite{vonKuk:2023jfd}
that it is given by a product of the total fragmentation probability $\chi_H$
and a perturbative matching coefficient $C_1$,
\begin{align} \label{eq:chi1h_ope}
 \chi_{1,H}(b_T, \mu, {\rho}) = C_1(b_T, \mu, {\rho}) \, \chi_H + \ord{\lqcd^2 b_T^2}
\,,\end{align}
where $C_1$ will be the target of our calculation in \sec{bhqet_tmd_fragmentation_factors_one_loop}.
Here we have also introduced the shorthand $\rho \equiv v^- = \sqrt{\zeta}/m$
for the boost of the hadron; in \refcite{vonKuk:2023jfd} we argued from consistency
with TMD factorization that $\chi_{1,H}$ develops an anomalous dependence on this variable
governed by the Collins-Soper kernel,
and we will make this expectation more explicit in \sec{bhqet_tmd_fragmentation_factors_renormalization}.

To evaluate $C_1$, we will use that -- for fully general $\lqcd \lesssim 1/b_T$ --
the unpolarized TMD fragmentation factor $\chi_{1,H}$ reduces to a vacuum matrix element of Wilson lines
when summed over all hadrons $H$ containing $Q$~\cite{vonKuk:2023jfd}.
Since $\sum_{H}\chi_{H} = 1$ and $C_1$ in \eq{chi1h_ope} is independent
of the hadronic state $H$,
it is then straightforward to obtain $C_1$
by evaluating the Wilson line correlator at perturbative $b_T$.

At the partonic level, \eqs{tmdffs_bhqet_all_order_final_results}{chi1h_ope} together imply
the following consistency condition for the partonic heavy-quark TMD FF
in the limit $m \gg k_T \sim 1/b_T$~\cite{vonKuk:2023jfd},
\begin{align} \label{eq:tmd_ff_unpol_consistency_heavy}
d_{1\,Q/Q}(z, b_T, \mu, \zeta)
= \delta(1-z) \,
C_m(m, \mu, {\zeta}) \,
C_1\Bigl(b_T, \mu, {\frac{\sqrt \zeta}{m}} \Bigr)
+ \ORd{\frac{1}{b_T m}}
\,,\end{align}
and in \sec{heavy_quark_limit_consistency} we will use this relation as a check of our one-loop results.
By contrast,
nonvalence (or disfavored) partonic heavy-quark TMD FFs become power-suppressed in the heavy-quark limit,
\begin{align}
i \neq Q \,: \quad d_{1\,Q/i}(z, b_T, \mu, \zeta) = \ORd{\frac{1}{b_T m}}
\,.\end{align}
The indirect effect ($D_{1\,h/i}$, $i \neq Q, \bar{Q}$) of the heavy quark
on light-hadron production for $m \gg k_T$
is leading in $1/m$.
It is governed by virtual contributions from collinear and soft so-called mass modes
and thus follows exactly the TMD PDF case~\cite{Pietrulewicz:2017gxc}.
Finally, the direct contribution ($D_{1\,h/Q}$) to light-hadron production
becomes strongly peaked at $z_h \to 0$ in the limit $m \gg k_T \gtrsim \lqcd$,
but for any finite value of $z_h$ (or argument $z$ of $\cJ_{j/Q}$ at the partonic level)
is kinematically suppressed as $m \gg k_T$;
we will verify this latter behavior in \sec{heavy_quark_limit_consistency},
but leave a dedicated analysis of its $z_h \to 0$ behavior
-- which is known to be subtle in fragmentation~\cite{Collins:2023cuo} --
to future work.

\subsection{Consistency conditions in the light-quark limit}
\label{sec:theory_small_mass}

In the case when the heavy-quark is light compared to $k_T$,
the matching at the scale $\mu \sim k_T$
takes exactly the standard form
for matching TMD FFs onto twist-2 collinear FFs~\cite{vonKuk:2023jfd},
\begin{align} \label{eq:light_tmd_ff_unpol_leading_twist}
D_{1\,\fh/Q}(z_\fh, b_T, \mu, \zeta)
= \frac{1}{z_\fh^2} \sum_i \int \! \frac{\df z}{z} \,
\cJ_{i/q}(z, b_T, \mu, \zeta) \,
D_{\fh/i}^{(n_\ell + 1)}\Bigl( \frac{z_\fh}{z}, \mu \Bigr)
+ \ord{m^2 b_T^2}
\,,\end{align}
where $\fh = h, H$ may be both a heavy or a light hadron
and $D_{\fh/j}^{(n_\ell + 1)}$ are collinear FFs
in a theory with $n_\ell$ light and one massive flavor,
i.e., the mass is the highest IR scale in the twist-2 matching.
Importantly, the TMD FF matching coefficients $\cJ_{i/q}$
are independent of these details in the IR
and are thus given by the universal TMD FF matching coefficients
in a theory with $n_f = n_\ell + 1$ \emph{light} degrees of freedom,
which are known to N$^3$LO~\cite{Luo:2020epw, Ebert:2020qef}.
The problem of the mass dependence is thus reduced
to the well-understood behavior of the collinear FFs,
which differs depending on whether $\fh$ is heavy or light~\cite{Mele:1990yq, Cacciari:2005ry},
\begin{align} \label{eq:coll_ff_flavor_decoupling}
D_{H/i}^{(n_\ell + 1)}(z_H, \mu)
&= d_{Q/i}(z_H, \mu) \,
\chi_H + \ORd{\frac{\lqcd}{m}}
\,, \nn \\
D_{h/i}^{(n_\ell + 1)}(z_h, \mu)
&= \sum_j \int \! \frac{\df z}{z} \, \cM_{j/i,T}(z, m, \mu) \,
D_{h/j}^{(n_\ell)}\Bigl( \frac{z_h}{z}, \mu \Bigr)
\,.\end{align}
Here $d_{Q/i}(z_H, \mu)$ is the partonic \emph{collinear} heavy-quark FF
and $\cM_{j/i,T}$ is the timelike matching function
governing the flavor decoupling in light-hadron collinear FFs.

Comparing \eqss{tmd_ff_unpol_bhqet}{nonvalence_matching_heavy_hadron}{nonvalence_matching_light_hadron}
to \eqs{light_tmd_ff_unpol_leading_twist}{coll_ff_flavor_decoupling}, we can read off the following leading-power
behavior of the relevant perturbative heavy-quark TMD matrix elements
in the limit $m \ll k_T \sim 1/b_T$~\cite{vonKuk:2023jfd},
\begin{align} \label{eq:consistency_relations_small_mass}
d_{1\,Q/i}(z, b_T, \mu, \zeta)
&= \frac{1}{z^2} \sum_j \int \! \frac{\df z'}{z'} \, \cJ_{j/i}(z', b_T, \mu, \zeta) \, d_{Q/j}\Bigl( \frac{z}{z'}, \mu \Bigr)
\,, \nn \\
\cJ_{k/i}(z, b_T, m, \mu, \zeta)
&= \frac{1}{z^2} \sum_j \int \! \frac{\df z'}{z'} \, \cJ_{j/i}(z', b_T, \mu, \zeta) \, \cM_{k/j,T}\Bigl( \frac{z}{z'}, m, \mu \Bigr)
\,.\end{align}
In \sec{light_quark_limit_consistency}, we will use these expressions, together with the various known
ingredients on the right-hand side, to perform cross checks
of our one-loop results in \sec{partonic_heavy_quark_tmd_ff} in all channels.

\section{Partonic heavy-quark TMD fragmentation at NLO}
\label{sec:partonic_heavy_quark_tmd_ff}

For ease of calculation and to make contact
with the splitting probability in \eq{heavy_quark_splitting_probability},
we will perform our calculation of the bare collinear matrix element
as well as its renormalization in momentum space,
i.e., as a function of $z$ and $k_T$,
and later compute various integral transforms
of the renormalized object directly.
Passing to momentum space,
the one-loop correction to the renormalized heavy-quark TMD FF
in \eq{def_d1qq_bT_ren} reads
\begin{align} \label{eq:d1qq_kT_ren_one_loop}
d_{1\,Q/Q}^{(1)}(z, k_T, \mu, \zeta)
&= \lim_{\eps \to 0} \biggl\{
   \delta(1-z) \, \frac{1}{\pi} \delta(k_T^2) \, Z_\UV^{(1)}(\mu, \zeta, \eps)
\\ & \quad
   +
\lim_{\eta \to 0} \Bigl[
   d_{1\,Q/Q}^{\bare\,(1)}(z, k_T, \eps, \eta, \zeta/\nu^2) + \delta(1-z) \, \tfrac{1}{2} S^{(1)}(k_T, \eps, \eta, \nu)
\Bigr]
\biggr\}
\nn \,,\end{align}
where we have used
that $d_{1\,Q/Q}^{(0)} = \delta(1-z) \, \frac{1}{\pi} \delta(k_T^2)$ at tree level.
(Here and in the following
we use the same symbol for functions of $b_T$ and their Fourier transforms,
as the meaning is always clear from the argument.)
Note that in our implementation of dimensional regularization in the transverse plane,
we make use of the azimuthal symmetry of the unpolarized TMD FF
and take the curly braces in \eq{d1qq_kT_ren_one_loop}
to be a density in $\pi k_T^2$ for convenience,
i.e., they have an integer mass dimension of $-2$,
where the factor of $\pi$
ensures that final results in $d = 4$ are properly normalized
azimuthally symmetric densities in vectorial $\vec{k}_T$~\cite{Jain:2011iu, Gaunt:2020xlc}.
With these preliminaries, the bare collinear matrix element in momentum space
is given by
\begin{align} \label{eq:def_d1qq_kT_bare}
d_{1\,Q/Q}^\bare(z, k_T, \eps, \eta, \zeta/\nu^2)
&=
\frac{1}{2 z^{1-2\eps} N_c}
\int \! \frac{\df \Omega_{2-2\eps}}{2\pi} \, k_T^{-2\eps}
\Tr \SumInt_{X}
\tr \Bigl[ \frac{\Sl{\bn}}{2} \,
\Mae{0}{W_\eta^\dagger \, \psi_Q}{Q X}
\\ & \quad
\times \Mae{Q X}{\bigl[
   \delta\bigl(p^-/z + \img\partial^-\bigr) \,
   \delta^{(2-2\eps)}\bigl(k_\perp^\mu + \img\partial_\perp^\mu\bigr) \,
   \bar\psi_Q \, W_\eta
\bigr]}{0}
\Bigr]
\nn \,.\end{align}
Here $\df \Omega_{2-2\eps}$ is the solid angle element for $k_\perp^\mu$
in $2-2\eps$ dimensions
and the $\delta$ functions involving derivative operators on the second line
act on the fields to their right,
i.e., they fix the total minus and perpendicular momentum injected into the compound operator.

Compared to a direct calculation in position space,
which involves expanding Bessel-like hypergeometric functions
in the dimensional regulator,
the main challenge in the momentum-space calculation
is the careful distributional treatment of the singularity structure
prior to expanding in $\eps$,
which we address in \sec{tmd_ff_reals_and_expansion}.
Renormalized results in momentum space are presented in \sec{tmd_ff_virtuals_and_uv_ren},
which we independently verify and further simplify
by moving to cumulant space in \sec{tmd_ff_cumulant_space}.
Results in $b_T$ space are then obtained by a straightforward
integral transform mapping finite functions onto functions,
and are given in \sec{tmd_ff_position_space}.

\subsection{Bare real-emission diagrams and distributional expansion}
\label{sec:tmd_ff_reals_and_expansion}

\begin{figure*}
\centering
\subfloat[]{
   \includegraphics[width=0.3\textwidth]{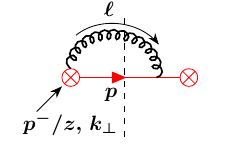}
}%
\subfloat[]{
   \includegraphics[width=0.3\textwidth]{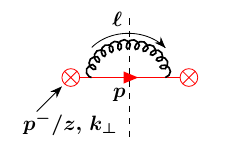}
}%
\subfloat[]{
   \includegraphics[width=0.3\textwidth]{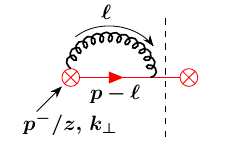}
}%
\caption{
Diagrams contributing to the partonic heavy-quark TMD FF
in Feynman gauge.
Heavy quark propagators and field insertions are indicated in red.
In diagrams (a) and (c) the gluon attaches
to the Wilson lines that are part
of the operators denoted by $\textcolor{red}{\otimes}$.
The dashed line indicates on-shell cuts.
Mirror diagrams for (a) and (c) are understood
and are included in expressions given in the text.
}
\label{fig:tmd_ff_diagrams}
\end{figure*}

The two contributing real-emission diagrams in Feynman gauge
are given in \fig{tmd_ff_diagrams}~(a) and (b),
where $X = g$ is a single gluon.
Including the mirror diagram for (a),
their contributions to $d_{1\,Q/Q}^{\bare\,(1)}$ evaluate to
\begin{align}
\label{eq:d1qq_diag_a}
d_{1\,Q/Q}^{\bare\,(a)}
&= \frac{\as C_F}{4\pi}  \frac{e^{\eps \gamma_E}}{\Gamma(1-\eps)}
\Bigl( \frac{\mu}{k_T} \Bigr)^{2\eps}
\Bigl( \frac{\sqrt{\zeta}}{\nu} \Bigr)^{-\eta}
\frac{1}{\pi z^{2-2\eps}}
\frac{z^\eta}{\zb^{1+\eta}} \frac{4z^3}{k_T^2 z^2 + m^2 \zb^2}
\,, \\ 
\label{eq:d1qq_diag_b}
d_{1\,Q/Q}^{\bare\,(b)}
&= \frac{\as C_F}{4\pi}  \frac{e^{\eps \gamma_E}}{\Gamma(1-\eps)} \,
\Bigl( \frac{\mu}{k_T} \Bigr)^{2\eps} \,
\frac{1}{\pi z^{2-2\eps}} \, 2 z^2 \zb
\nn \\ & \quad \times
\Bigl[
\frac{
   k_T^2 z^2 + m^2 (1 - 4z + z^2)
   - \eps (k_T^2 z^2 + m^2 \zb^2)
}{
   (k_T^2 z^2 + m^2 \zb^2)^2
}
\Bigr]
\,,\end{align}
where we defined the shorthand $\zb \equiv 1-z$
and $\as = \as(\mu)$ is the renormalized $\MSbar$ coupling.
Note that to our working order, all results in this paper are independent
of whether they are expressed in terms of $\as^{(n_f)}$ or $\as^{(n_\ell)}$.
We suppress overall factors of $\Theta(z)$, $\Theta(\zb)$, or $\Theta(k_T^2)$
from the final-state on-shell conditions in the following,
but stress that they are formally present in all the distributional identities
we use or derive, and also multiply all finite remainder terms.

The two seemingly simple expressions above feature an intricate
interplay of singularities as $k_T \to 0$ and/or $z \to 1$ ($\zb \to 0$),
which in particular arise as the quark propagator goes on shell
for $k_T^2 z^2 + m^2 \zb^2 \to 0$.
At this step, the singularities are regulated
by powers of $\zb^{-\eta}$, dimensional regularization, or both.
However, if we restrict to finite $z < 1$ and $k_T > 0$,
the limits $\eta \to 0$ and $\eps \to 0$ can be taken right away
and all other contributions in the renormalized one-loop formula
including the virtual diagrams drop out in \eq{d1qq_kT_ren_one_loop},
leaving behind a finite result from the sum of (a) and (b),
\begin{align} \label{eq:d1QQ_kT_ren_one_loop_bulk}
z < 1 ~ \text{and} ~ k_T > 0 \,: \quad
d_{1\,Q/Q}^{(1)}(z, k_T, \mu, \zeta)
= \frac{\as C_F}{4\pi} \frac{2}{\pi}\frac{k_T^2 z^2 (1 + z^2) + m^2 \zb^4}{\zb (k_T^2 z^2 + m^2 \zb^2)^2}
\,.\end{align}
This is, of course, exactly the differential splitting probability
for $Q \bar{Q} \to (Qg) \bar{Q}$ in the quasi-collinear limit
that we considered in \eq{heavy_quark_splitting_probability},
but now the field-theoretic definition of $d_{1\,Q/Q}$
provides us with explicit regulators controlling the singular limits.
It is in this sense that the heavy-quark TMD FF,
after including the virtual diagram and performing the renormalization,
will provide a fully differential extension
of the splitting probability into the unresolved limit(s)
that is finite in $d = 4$, universal, and embedded in factorization theorems.

To proceed, we first expand the Wilson line diagram~(a)
in the rapidity regulator $\eta$, using the standard identity
\begin{align} \label{eq:expanding_in_plus_distros_eta}
\frac{z^\eta}{\zb^{1+\eta}}
= - \frac{\delta(\zb)}{\eta} + \cL_0(\zb) + \ord{\eta}
\,.\end{align}
This expansion is valid here because the remaining terms in \eq{d1qq_diag_a}
are still dimensionally regulated at $\zb = 0$ as $k_T \to 0$
i.e., the remainder of the expression is a good test function
for finite $\eps \neq 0$.
Since the mass enters diagram~(a) only through the combination $m \zb$,
the action of the $\delta(\zb)/\eta$ term
on the rest of the diagram is to reduce it to the massless limit.
This in particular means that the coefficient of the $1/\eta$ pole
(including its exact $\eps$ dependence)
is equal to the one for the massless TMD FF,
as it must for it to cancel the pole in the $\eta$-regulated soft function,
which is independent of the mass at this order.
For our implementation of dimensional regularization, the soft function is given by
\begin{align} \label{eq:bare_tmd_soft_function}
\frac{1}{2} S^{(1)}(k_T, \eps, \eta, \nu)
= \frac{\as C_F}{4\pi} \frac{e^{\eps \gamma_E}}{\Gamma(1-\eps)}
\Bigl( \frac{\mu}{k_T} \Bigr)^{2\eps}
\frac{1}{\pi k_T^2} \Bigl[ + \frac{4}{\eta} + 2 \ln \frac{\nu^2}{k_T^2} + \ord{\eta} \Bigr]
\,.\end{align}
Combining this with diagram~(a), the $\eta \to 0$ limit can be taken,
and we find
\begin{align} \label{eq:d1QQ_a_s}
d_{1\,Q/Q}^{\bare\,(a + s)}
&\equiv
\lim_{\eta \to 0} \Bigl[
   d_{1\,Q/Q}^{\bare\,(a)}(z, k_T, \eps, \eta, \zeta/\nu^2) + \delta(1-z) \, \tfrac{1}{2} S^{(1)}(k_T, \eps, \eta, \nu)
\Bigr]
\\
&= \frac{\as C_F}{4\pi} \frac{e^{\eps \gamma_E}}{\Gamma(1-\eps)}
\frac{1}{\pi z^{2-2\eps}} \biggl\{
   \frac{\mu^{2\eps}}{k_T^{2+2\eps}} \Bigl[
      2 \ln \frac{\zeta}{k_T^2} \delta(\zb)
      + 4\cL_0(\zb)
   \Bigr]
   + \frac{\mu^{2\eps}}{m^{2+2\eps}} \, f^{(a)}(k_T^2/m^2,z,\eps)
\nn \biggr\}
\,.\end{align}
Here the first term in curly braces,
which is left behind from the cancellation against the soft function
and the same as for the massless case,
is straightforward to expand in $\eps$
using identities like \eqs{expanding_in_plus_distros}{expanding_in_plus_distros_with_log}.
This produces, among others, a term $\tfrac{1}{\eps^2} \delta(\zb) \delta(k_T^2)$
as well as a term $\tfrac{1}{\eps} \cL_0(\zb) \delta(k_T^2)$.
The nontrivial part is to expand the dimensionless function
\begin{align} \label{eq:def_f_a}
f^{(a)}(x, z, \eps) \equiv x^{-1 - \eps} \frac{-4 + 4 z - 4 x z^2}{\zb^2 + x z^2}
\end{align}
in a Laurent series in $\eps$ involving plus distributions of $x$ and $z$.
To do so, we recall that plus distributions in one variable $x$
formally arise in perturbative calculations as
\begin{align} \label{eq:plus_distros_1d_why}
f(x, \eps) = \bigl[f(x, \eps) \bigr]_+ + \delta(x) \, F(\eps)
\,, \qquad
F(\eps) = \int_0^1 \! \df x' \, f(x', \eps)
\,,\end{align}
where $f(x, \eps) \sim x^{-1-\eps} + \ord{x^0}$
has a dimensionally regulated singular limit $x \to 0$
on the first left-hand side.
By contrast, on the right-hand side $f(x, \eps)$
may be expanded in $\eps$ since each term
in the expansion is now regulated by the plus prescription,
whereas the explicit poles are typically isolated
in the Laurent series of the total integral $F(\eps)$.
Of course, this is precisely how relations like \eq{expanding_in_plus_distros_eta} are derived.

The identity in \eq{plus_distros_1d_why} generalizes
to our case of two variables $x$ and $z$, cf.~\refcite{Lustermans:2019cau},
\begin{align} \label{eq:plus_distros_2d_how}
f(x, z, \eps)
= \bigl[f(x, z, \eps) \bigr]_{+,+}
+ \delta(x) \bigl[ F_x(z, \eps) \bigr]_+
+ \delta(\zb) \bigl[ F_z(x, \eps) \bigr]_+
+ \delta(x) \, \delta(\zb) \, F_{xz}(\eps)
\,,\end{align}
where
\begin{align} \label{eq:plus_distros_2d_fixing_unknowns_in_ansatz}
F_x(z, \eps) &\equiv \int_0^1 \! \df z' \, f(x, z', \eps)
\,, \qquad
F_z(x, \eps) \equiv \int_0^1 \! \df x' \, f(x', z, \eps)
\,, \nn \\ & \qquad
F_{xz}(\eps) \equiv \int_0^1 \! \df x \, \int_0^1 \! \df z' \, f(x, z', \eps)
\end{align}
and the two-dimensional plus bracket $[\dots]_{+,+}$ is defined through
its action on test functions $g(x, z)$ as
\begin{align} \label{eq:def_plus_plus}
&\int_0^1 \! \df x \int_{0}^{1} \! \df z \, \bigl[f(x, z, \eps) \bigr]_{+,+} \, g(x, z)
\nn \\
&\equiv \int_0^1 \! \df x \int_{0}^{1} \! \df z \, f(x, z, \eps) \, \bigl[
   g(x, z)
   - g(0, z)
   - g(x, 0)
   + g(0, 0)
\bigr]
\,.\end{align}
Note that in our case with $x = k_T^2/m^2$,
the integration domain can also extend to $x_* > x > 1$,
in which case the $[\dots]_{+,+}$ bracket reduces to a one-dimensional one,
\begin{align}
\int_1^{x_*} \! \df x \int_{0}^{1} \! \df z \, \bigl[f(x, z, \eps) \bigr]_{+,+} \, g(x, z)
\equiv \int_1^{x_*} \! \df x \int_{0}^{1} \! \df z \, f(x, z, \eps) \, \bigl[
   g(x, z)
   - g(x, 0)
\bigr]
\,,\end{align}
while integration domains with other shapes
can be implemented through suitable test functions.
So far, \eq{plus_distros_2d_how} together with \eq{def_plus_plus} is a tautology;
however, as in the one-dimensional case in \eq{plus_distros_1d_why},
the crucial step is that the expansion in $\eps$ may now be performed
under each of the various plus brackets.

For the case of $f = f^{(a)}$ as defined in \eq{def_f_a},
the various integrals in \eq{plus_distros_2d_fixing_unknowns_in_ansatz}
are all straightforward to evaluate for general $\eps$
in terms of incomplete Beta and digamma functions.
If we restrict to their expansions in $\eps$ for brevity,
\eq{plus_distros_2d_how} evaluates to
\begin{align} \label{eq:eps_expansion_fa}
f^{(a)}(x, z, \eps)
&= \bigl[ f^{(a)}(x, z, 0) \bigr]_{+, +}
+ \delta(x) \biggl\{
   \frac{4}{\eps} \cL_0(\zb)
   + 4 \biggl[ \frac{4z \ln \bigl( 1 + z^2/\zb^2\bigr)}{\zb} \biggr]_+
\biggr\}
\nn \\ & \quad
+ \delta(\zb) \biggl[
   \frac{2 (1 - x) \ln x - 4(\pi \sqrt{x} + x + x^2)}{x(1+x)^2}
\biggr]_+
\nn \\ & \quad
+ \delta(\zb) \delta(x) \biggl\{ - \frac{2}{\eps^2} - 2 \pi - \frac{5\pi^2}{6} \biggr\}
+ \ord{\eps}
\,.\end{align}
In this form, the double pole $1/\eps^2$ is isolated at $\zb = x = 0$.
The term proportional to $\tfrac{1}{\eps} \cL_0(\zb)$ along the $x = 0$ boundary
ends up canceling a corresponding term from the $\eps$ expansion
of the first term in curly braces on the second line of \eq{d1QQ_a_s},
such that $d_{1\,Q/Q}^{\bare\,(a + s)}$ is finite
as $\eps \to 0$ for $z < 1$ for any $x \geq 0$.
The double pole $\tfrac{1}{\eps^2} \delta(\zb) \delta(k_T^2)$ similarly cancels.

The second diagram in \fig{tmd_ff_diagrams} does not require rapidity regularization.
In order to expand it in the dimensional regulator,
we repeat the above procedure,
where we define
\begin{align}
d_{1\,Q/Q}^{\bare\,(b)}
&\equiv \frac{\as C_F}{4\pi} \frac{e^{\eps \gamma_E}}{\Gamma(1-\eps)}
\frac{1}{\pi z^{2-2\eps}}
   \frac{\mu^{2\eps}}{m^{2+2\eps}} \, f^{(b)}(k_T^2/m^2,z,\eps)
\,, \nn \\
f^{(b)}(x, z,\eps)
&= 2 z^2 \zb \, x^{-2\eps}
\frac{
   x z^2 + (1 - 4z + z^2)
   - \eps (x z^2 + \zb^2)
}{
   (x z^2 + \zb^2)^2
}
\,.\end{align}
The integrals in \eq{plus_distros_2d_fixing_unknowns_in_ansatz} for $f = f^{(b)}$
can again be performed including their exact dependence on $\eps$,
leading to digamma and $_2F_1$ hypergeometric functions.
Expanding the latter in $\eps$ using the \texttt{HypExp 2.0} package~\cite{Huber:2005yg},
we find the following distributional expansion of the dimensionless function $f^{(b)}$,
\begin{align} \label{eq:eps_expansion_fb}
f^{(b)}(x, z,\eps)
&= \bigl[ f^{(b)}(x, z, 0) \bigr]_{+, +}
+ \delta(x) \biggl[
   \frac{4 z^3}{\zb (2 \zb z - 1)} + 2 \zb \ln \Bigl (1 + \frac{z^2}{\zb^2} \Bigr)
\biggr]_+
\\ & \quad
+ \delta(\zb) \biggl[
   \frac{x (5+x) \ln x + (-1+\pi \sqrt{x} -x) (2 - 3x - 3x^2)}{x (1+x)^3}
\biggr]_+
\\ & \quad
+ \delta(\zb) \delta(x) \biggl\{
\frac{2}{\eps} + 1 + \frac{3 \pi}{2}
\biggr\}
+ \ord{\eps}
\,.\end{align}
Combining \eqs{eps_expansion_fa}{eps_expansion_fb}
produces an explicit result for the sum of real-emission diagrams
expanded in $\eps$.
The finite terms involving the one and two-dimensional plus distributions
are lengthy and reappear in our renormalized result below,
so we only quote the poles for reference:
\begin{align}
d_{1\,Q/Q}^{\bare\,(a + s)} + d_{1\,Q/Q}^{\bare\,(b)}
= \frac{\as C_F}{4\pi} \frac{1}{\pi z^2} \, \biggl\{
   \delta(\zb) \delta(k_T^2) \Bigl(
      - \frac{2}{\eps} \ln \frac{\zeta}{m^2} + \frac{2}{\eps}
   \Bigr)
   + \ord{\eps^0}
\biggr\}
\,.\end{align}

\subsection{Virtual contributions and UV renormalization}
\label{sec:tmd_ff_virtuals_and_uv_ren}

The virtual diagram~(c) from \fig{tmd_ff_diagrams}
is common to any collinear matrix element involving massive fermions
and can be evaluated in a straightforward way
using integration by contours for the $\ell^+$ component of the loop momentum $\ell$,
Feynman parametrization for the $\ell_\perp$ integral,
and an analytic $\ell^-$ integral at the end.
Including the mirror diagram,
the result reads
\begin{align}
d_{1\,Q/Q}^{\bare\,(c)}
= \frac{\as C_F}{4\pi} \, \delta(1-z) \frac{1}{\pi} \delta(k_T^2) \,
\frac{2 e^{\eps \gamma_E} \Gamma(\eps)}{\eps(1 - 2\eps)} \frac{\mu^{2\eps}}{m^{2\eps}}
\,,\end{align}
and contributes a double $1/\eps^2$ pole.
So far our calculation was performed in terms of bare quark fields,
and we still have to account for \eq{quark_field_ren}.
Using the standard on-shell renormalization factor
for a quark field with pole mass $m$ in dimensional regularization,
\begin{align}
Z_{\psi,\mathrm{OS}} \equiv 1 + Z_{\psi,\mathrm{OS}}^{(1)} + \ord{\as^2}
\,, \qquad
Z_{\psi,\mathrm{OS}}^{(1)} = - \,\frac{\as C_{F}}{4 \pi}
\Big( \frac{3}{\eps} + 3 \ln \frac{\mu^2}{m^2} + 4\Big)
\end{align}
the rapidity-renormalized collinear correlator
on the second line of \eq{d1qq_kT_ren_one_loop}
is given by
\begin{align} \label{eq:d1qq_final_bare_result}
&\lim_{\eta \to 0} \Bigl[
   d_{1\,Q/Q}^{\bare\,(1)}(z, k_T, \eps, \eta, \zeta/\nu^2) + \delta(1-z) \, \tfrac{1}{2} S^{(1)}(k_T, \eps, \eta, \nu)
\Bigr]
\nn \\
&= d_{1\,Q/Q}^{\bare\,(a + s)} + d_{1\,Q/Q}^{\bare\,(b)}
+ d_{1\,Q/Q}^{\bare\,(c)} + \delta(1-z) \frac{1}{\pi} \delta(k_T^2) Z_{\psi}^{(1)}
\nn \\
&= \frac{\as C_F}{4\pi} \frac{1}{\pi z^2} \, \biggl\{
   \delta(\zb) \delta(k_T^2) \Bigl(
      \frac{2}{\eps^2} + \frac{3}{\eps}
      - \frac{2}{\eps} \ln \frac{\zeta}{\mu^2}
   \Bigr)
   + \ord{\eps^0}
\biggr\}
\,,\end{align}
and we continue to only quote the poles for brevity.
The $\MSbar$ UV renormalization factor for quark TMD PDFs and FFs is likewise well known
and independent of the mass~\cite{Boussarie:2023izj},
\begin{align} \label{eq:Z_uv_one_loop}
Z_\UV^{(1)}(\mu, \zeta, \eps)
= - \frac{\as C_F}{4\pi}
\biggl\{
\frac{2}{\eps^2} + \frac{3}{\eps} - \frac{2}{\eps} \ln \frac{\zeta}{\mu^2}
\biggr\}
\,.\end{align}
It precisely cancels the poles from \eq{d1qq_final_bare_result}
when inserting both into \eq{d1qq_kT_ren_one_loop}, as expected.
We thus confirm by an explicit one-loop calculation
that the TMD FF for heavy quarks obeys standard TMD evolution equations,
with secondary mass effects in the Collins-Soper kernel
at higher loop orders understood~\cite{Pietrulewicz:2017gxc}.

Taking the $\eps \to 0$ limit, we obtain our final result
for the $\ord{\as}$ correction to the renormalized partonic heavy-quark TMD FF:
\begin{align} \label{eq:d1qq_kT_ren_one_loop_result}
&d_{1\,Q/Q}^{(1)}(z, k_T, \mu, \zeta)
\nn \\
&= \frac{\as C_F}{4\pi} \frac{1}{\pi z^2} \, \biggl\{
   \delta(\zb) \Bigl[
      2 \ln \frac{\zeta}{\mu^2} \cL_0(k_T^2, \mu^2)
      + 2 \ln \frac{\mu^2}{m^2} \cL_0(k_T^2, m^2)
      - \ln^2 \frac{\mu^2}{m^2} \delta(k_T^2) + 3 \ln \frac{\mu^2}{m^2} \delta(k_T^2)
   \Bigr]
\nn \\ & \qquad
+ \frac{1}{m^2} \biggl[
   \frac{2 x z^4 (1 + z^2) + 2 z^2 \zb^4}{\zb (x z^2 + \zb^2)^2}
\biggr]_{+,+}
+ \delta(k_T^2) \biggl[
   \frac{2 (1 + z^2)}{\zb}  \ln \Bigl(1 + \frac{z^2}{\zb^2}\Bigr)
   - \frac{4 z^3}{\zb (1 - 2 \zb z)}
\biggr]_{+}
\nn \\ & \qquad
+ \delta(\zb) \, \frac{1}{m^2} \biggl[
   -\frac{2 + 3 x + 4 x^2 + 3 x^3 +\pi \sqrt{x} (2 + 7x + x^2) + x (1 + 7x + 2x^2) \ln x}{x (1 + x)^3}
\biggr]_{+}
\nn \\ & \qquad
+\delta(\zb) \, \delta(k_T^2) \Bigl( 5 - \frac{\pi}{2} - \frac{2 \pi^2}{3} \Bigr)
\biggr\}
\,.\end{align}
Here we still use the dimensionless variable $x \equiv k_T^2/m^2$.
To simplify the result we have combined terms under $+,+$ and $+$ brackets
as much as possible, exploiting their linearity,
reexpressed plus distributions $\cL_n(k_T^2, \mu^2)$
in terms of $\cL_n(k_T^2, m^2)$ and logarithms of $\mu^2/m^2$,
and used that products of one-dimensional plus distributions
can be written as (trivial) two-dimensional ones, i.e.,
\begin{align}
\cL_{0}(\zb) \, \cL_0(k_T^2, m^2)
= \cL_{0}(\zb) \, \frac{1}{m^2} \cL_0(k_T^2/m^2)
= \frac{1}{m^2} \Bigl[ \frac{1}{x} \frac{1}{\zb} \Bigr]_{+,+}
\,.\end{align}
\Eq{d1qq_kT_ren_one_loop_result} is the main result of this section.
Up to the terms on the first line of the right-hand side,
which are predicted by the RGE and vanish for $\zeta = \mu^2 = m^2$,
we have cast the heavy-quark TMD FF precisely in the form of \eq{plus_distros_2d_how},
i.e., as a two-dimensional plus distribution
with nontrivial functions of $z$ and $k_T$
living on the respective boundaries
and an overall boundary contribution at $z = 1$ and $k_T = 0$.
This structure was already illustrated in \fig{3d_plot_d1qq},
where the partial integrals over one of the variables
can immediately be read off from the respective boundary terms.
Of course, the total contribution in the bulk at $z < 1$ and $k_T > 0$,
which can simply be read off from the content of the $[\dots]_{+,+}$ brackets
accounting for the prefactor of $1/z^2$,
is still equal to the simple result we found in \eq{d1QQ_kT_ren_one_loop_bulk}
and thus to the quasi-collinear splitting probability in \eq{heavy_quark_splitting_probability}.

\subsection{Results in cumulant space}
\label{sec:tmd_ff_cumulant_space}

Fixed-order subtraction methods,
as well as certain $q_T$ resummation formalisms
like the \texttt{RadISH} approach~\cite{Bizon:2017rah},
require the cumulative distribution in $k_T$ integrated
over transverse momenta $\vec{k}_T$ with $\abs{\vec{k}_T} \leq \ktcut$ as an input,
\begin{align}
d_{1\,Q/Q}(z, \ktcut, \mu, \zeta)
\equiv \pi \! \int^{(\ktcut)^2} \! \df (k_T^2) \, d_{1\,Q/Q}(z, k_T, \mu, \zeta)
\end{align}
often simply referred to as the cumulant.
As for the Fourier transform we will indicate the cumulative distribution
simply through its argument $\ktcut$.

The organization of our final result in \eq{d1qq_kT_ren_one_loop_result}
in terms of plus distributions defined on rectangular domains in $(z, k_T^2)$
makes it very straightforward to compute the cumulative distribution
because the cumulant constitutes a special case of \eq{def_plus_plus} for a unit test function.
Specifically, for plus-bracketed functions $f$
of either a single variable $x = k_T^2/m^2$
or of both $x$ and $z$,
we have the following integrals up to $\xcut \equiv (\ktcut/m)^2$:
\begin{align} \label{eq:plus_distros_2d_how_to_take_x_cumulant}
\int^{\xcut} \! \df x \, \bigl[ f(x, z) \bigr]_{+,+}
&= \biggl[ \int_1^{\xcut} \! f(x, z) \biggr]_+
\,, \nn \\
\int^{\xcut} \! \df x \, \bigl[ f(x) \bigr]_+
&= \int_1^{\xcut} \! f(x)
\,,\end{align}
where the remaining one-dimensional plus bracket on the right-hand side of the first line
refers to the $z$ dependence of the integral inside it,
while the latter's $\xcut$ dependence is function valued.

The integrals in \eq{plus_distros_2d_how_to_take_x_cumulant}
are straightforward to evaluate
for the respective terms in \eq{d1qq_kT_ren_one_loop_result}.
For purposes of our presentation, we find it convenient
to further decompose the $z$ plus distribution arising
from the cumulant of the $[\dots]_{+,+}$ term as
\begin{align} \label{eq:s_r_decomposition}
\bigl[ s(\xcut, z) + r(\xcut, z) \bigr]_+
= \bigl[ s(\xcut, z) \bigr] + r(\xcut, z)
- \delta(\zb) \int_0^1 \! \df z' \, r(\xcut, z')
\,,\end{align}
where $s(\xcut, z)$ is defined as the leading term
under the brackets for $\zb \ll \xcut \sim 1$,
with homogeneous $\sim 1/\zb$ scaling,
while $r(\xcut, z)$ is defined as the remainder
that is at most logarithmically divergent as $z \to 1$.
Notably, we find that in this operation, the $\xcut$ dependence of $s(z)$
is purely logarithmic, while the power-like $\xcut$ dependence
of the integrated $r(\xcut, z')$ multiplying $\delta(\zb)$
precisely compensates that of the complicated cumulant integral
of the one-dimensional plus distribution in $x$ in \eq{d1qq_kT_ren_one_loop_result}.

Collecting terms, our final result for the NLO correction
to the renormalized heavy-quark TMD FF
in cumulant space reads
\begin{align} \label{eq:d1QQ_kTcut}
&d_{1\,Q/Q}^{\,(1)}(z, \ktcut, \mu, \zeta)
\nn \\[0.2em]
&= \frac{\as C_F}{4\pi} \frac{1}{z^2}
\bigg\{  \delta(1-z) \bigg [
- 4 \ln \frac{\mu}{\ktcut} \ln\frac{\zeta}{m^2}
+ 4 \ln^{2}\frac{\mu}{m}
+ 6 \ln \frac{\mu}{m}
- 4 \ln^2 \frac{\ktcut}{m}
+ 4 - \frac{\pi^{2}}{6}
\bigg]
\nn \\[0.2em]
& \qquad \qquad \qquad
- 4 \Bigl(1-2 \ln \frac{\ktcut}{m} \Bigr) \, \cL_{0}(1-z)
- 8 \cL_{1}(1-z)
+ \cR\Bigl[z, \Bigl(\frac{\ktcut}{m} \Bigr)^{\!2\,} \Bigr]
\bigg\}
\,,\end{align}
where we defined
\begin{align} \label{eq:def_cR}
\cR(z,\xcut) \equiv & \,\,
\frac{4-4z(1-\xcut z)}{\zb^2 + \xcut z^2} + 4 \ln \frac{\zb^2}{\xcut}
+ \frac{1}{\zb}\Bigl[
   2(1+z^2) \ln \Bigl( 1 + \frac{\xcut z^2}{\zb^2} \Bigr)
   - 4 z \ln \frac{\xcut}{\zb^2} \Bigr]
\,.\end{align}
We note that $\cR(z, \xcut) = \ord{\zb^0}$ for $z \to 1$ and also $\cR(z, \xcut) = \ord{z^0}$
for $z \to 0$, so it is integrable in all limits.
For later reference we note its total integral,
\begin{align} \label{eq:cR_total_z_integral}
\int_0^1 \! \df z \, \cR(z, \xcut)
&= -4 \mathrm{Li}_2\Bigl(1 + \frac{1}{\img \sqrt{\xcut}}\Bigr)-4 \mathrm{Li}_2\Bigl(1 - \frac{1}{\img \sqrt{\xcut}}\Bigr)
\nn \\ & \quad
- \frac{3 \xcut^2 + 3 \xcut + 2}{(1 + \xcut)^2} \ln \xcut
+ \frac{2 \xcut (\xcut - \pi \sqrt{\xcut} + 1)}{(1 + \xcut)^2}
\,.\end{align}
To check \eq{d1QQ_kTcut}, we have performed an independent calculation
of the bare collinear correlator in dimensional regularization
and renormalized it directly in cumulant space,
This approach
involves more complicated hypergeometric functions of combinations of $\xcut$ and $z$
in intermediate steps;
in particular, the challenge in this case is that their expansion in $\eps$ produces
terms that are singular as $z \to 1$, even after subtracting terms
that scale as $1/\zb^{1 + \eps}$ for general $\eps$.
This means that the full hypergeometric result in dimensional regularization
needs to be treated as in \eq{plus_distros_1d_why} before expanding in $\eps$.
The final result is in full agreement with \eq{d1QQ_kTcut}.

\Eq{d1QQ_kTcut} has the -- at first sight -- remarkable property
that the dependence of the coefficient of $\delta(1-z)$ on $\ktcut/m$
is purely logarithmic, while a power-like dependence on $\ktcut/m$
would in general be allowed by our power counting assumption $\ktcut \sim m$
for the heavy-quark TMD FF.
Instead, after the cancellations discussed below \eq{s_r_decomposition} have taken place,
the nontrivial power-like dependence on $\ktcut/m$
arises solely in the $\cR(z, \xcut)$ term, which is regular as $z \to 1$.
In \sec{joint_res} we will explain this fact from the perspective
of joint threshold and $k_T$ factorization,
and give all-order expressions for the $z \to 1$ limit of $d_{1\,Q/Q}$
at fixed $k_T / m \sim b_T m \sim 1$.

\subsection{Results in position space}
\label{sec:tmd_ff_position_space}

We now move on to position-space results in terms of the $\vec{b}_T$
variable Fourier-conjugate to $\vec{k}_T$,
which are the key input for producing resummed predictions
from solving multiplicative $b_T$-space RGEs
and Fourier-transforming the final result.
As is standard, the two-dimensional Fourier transform
of the azimuthally symmetric renormalized TMD FF at hand
is simply given by a Fourier-Bessel (or Hankel) transform of zeroth order,
\begin{align} \label{eq:hankel_transform_kT_to_bT}
d_{1\,Q/Q}(z, b_T, \mu, \zeta)
= 2\pi \int_0^\infty \! \df k_T \, k_T \, J_0(b_T k_T) \, d_{1\,Q/Q}(z, k_T, \mu, \zeta)
\,,\end{align}
where $J_n$ is the $n$th-order Bessel function of the first kind.
It is convenient to integrate the above relation by parts,
\begin{align} \label{eq:hankel_transform_kTcut_to_bT}
d_{1\,Q/Q}(z, b_T, \mu, \zeta)
= b_T \int_0^\infty \! \df \ktcut \, J_1(b_T\ktcut) \, d_{1\,Q/Q}(z, \ktcut, \mu, \zeta)
\,.\end{align}
This lets us compute the Fourier-space result directly
from the cumulant in \eq{d1QQ_kTcut} obtained in the previous section,
which already had function-valued (and simpler) $\ktcut$ dependence,
cf.\ the more complicated original momentum-space result in \eq{d1qq_kT_ren_one_loop_result}.

Evaluating the relevant Hankel transforms,
our result for the NLO correction to the renormalized heavy-quark TMD FF
in position space reads
\begin{align} \label{eq:d1QQ_bT}
d_{1\,Q/Q}^{\,(1)}(z, b_T, \mu, \zeta)
&= \frac{\as C_F}{4\pi} \frac{1}{z^2}
\bigg\{  \delta(1-z) \bigg [
- 2 L_b \ln\frac{\zeta}{m^2}
+ 4 \ln^{2}\frac{\mu}{m}
+ 6 \ln \frac{\mu}{m}
- L_y^2
+ 4 - \frac{\pi^{2} }{6}
\bigg]
\nn \\[0.2em]
& \qquad \qquad
- 4 (1 + L_y ) \, \cL_{0}(1-z)
- 8 \cL_{1}(1-z)
+ \tcR(z, b_T m)
\bigg\}
\,,\end{align}
where we have used the shorthands
\begin{align} \label{eq:def_Lb_y_Ly}
L_b &\equiv 2\ln \frac{b_T \mu}{2 e^{-\gamma_E}}
\,, \qquad
y \equiv b_T m
\,, \quad
L_y \equiv 2\ln \frac{y}{2 e^{-\gamma_{E}}}
\,, \end{align}
and the regular term $\tcR(y)$ is again integrable for $z \to 0,1$
and given by
\begin{align} \label{eq:def_tcR}
\tcR(z, y)
&\equiv y \int_{0}^{\infty} \! \df t \,  J_{1}(t y) \, \cR (z, t^2)
\nn \\
&= \frac{4}{\zb} \Bigl[
   1 + L_y + (1+z^2) K_0\Bigl( \frac{y \zb}{z} \Bigr)
   - y \zb K_1\Bigl( \frac{y \zb}{z} \Bigr) + 2\ln \zb
\Bigr]
\,.\end{align}
While we were not able to find a closed-form expression
for the total $z$ integral of $\tcR$,
it can readily be obtained numerically
or analytically in asymptotic limits
through the Hankel transform of \eq{cR_total_z_integral}.

\subsection{Nonvalence channels at \texorpdfstring{$\ord{\as}$}{O(as)}}
\label{sec:tmd_ff_other_channels}

\begin{figure*}
\centering
\subfloat[]{
   \includegraphics[width=0.3\textwidth]{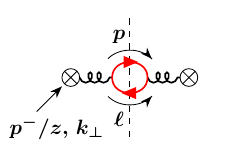}
}%
\subfloat[]{
   \includegraphics[width=0.3\textwidth]{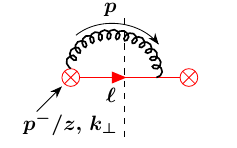}
}%
\subfloat[]{
   \includegraphics[width=0.3\textwidth]{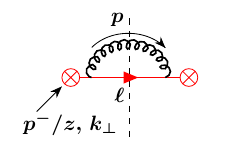}
}%
\caption{
Diagrams contributing to the gluon-initiated partonic heavy-quark TMD FF
and the mass-dependent TMD FF matching coefficients at $\ord{\as}$.
Heavy quark propagators and field insertions are indicated in red.
The mirror diagrams for (a) and (b) are understood
and are included in expressions given in the text.
}
\label{fig:tmd_ff_diagrams_offdiagonal}
\end{figure*}

For completeness we also compute the partonic TMD FFs
(or TMD FF matching coefficients onto collinear distributions)
for all remaining partonic channels involving heavy quarks.
The relevant diagrams at $\ord{\as}$ are given in \fig{tmd_ff_diagrams_offdiagonal},
where we labeled the identified final-state ``hadron'' by the momentum $p$.
Note that for diagrams~(b) and (c)
compared to the otherwise identical diagrams in \fig{tmd_ff_diagrams},
this also changes the definition of $k_\perp$,
as the coordinate system needs to be adjusted
such that the gluon now has vanishing transverse momentum.
Results for antiquarks are identical at this order.
For the $g \to Q$ contribution, diagram~(a) evaluates to
\begin{align} \label{eq:d1Qg_kT_ren_one_loop}
d_{1\,Q/g}^{(1)}(z, k_T)
=\frac{\as}{4 \pi} T_F\frac{2}{\pi z^2} \frac{k_T^2 z^4 \bigl(2 z^2-2 z+1\bigr)+m^2 z^2}{\bigl(k_T^2 z^2+m^2\bigr)^2}
\,.\end{align}
For gluon production off the heavy quark,
we find after summing over diagrams~(b) and (c)
and including the mirror diagram for~(b),
\begin{align} \label{eq:cJ1gQ_kT_ren_one_loop}
\cJ_{g/Q}^{(1)}(z, k_T, m)
=\frac{\as}{4 \pi} C_F  \frac{2}{\pi }\frac{k_T^2 \bigl(z^2-2 z+2\bigr)+m^2 z^2}{ z \bigl(k_T^2+m^2\bigr)^2}
\,.\end{align}
It is straightforward to verify from \eq{d1QQ_kT_ren_one_loop_bulk}
that at $0< z < 1$ and $k_T > 0$ the above expression satisfies
$1/z^2 \cJ_{g/Q}^{(1)}(z, k_T, m) = d_{1Q/Q}^{(1)}(1-z, k_T\frac{z}{1-z})$,
as expected from $z \leftrightarrow 1-z$ and the simultaneous
change of coordinate system;
similarly, $d_{1\,Q/g}^{(1)}(z, k_T) = d_{1Q/g}^{(1)}(1-z, k_T\frac{z}{1-z})$.
Note that the bare results in \eqs{d1Qg_kT_ren_one_loop}{cJ1gQ_kT_ren_one_loop}
are finite at $z > 0$ in $d = 4$ and without a rapidity regulator on their own,
as expected because there is no cross term with the soft function
or the UV renormalization at this order.
This is in particular the case as $k_T \to 0$, where all singularities in $k_T$
are now cut off by the mass for any value of $z > 0$.

The effect of the heavy quark on the gluon-onto-gluon
matching coefficient at one loop can be written as
\begin{align}
\cJ_{g/g}^{(1)}(z, k_T, m, \mu, \zeta)
\equiv \cJ_{g/g}^{(1,n_\ell)}(z, k_T, \mu, \zeta)
+ \Delta \cJ_{g/g}^{(1,h)}(z, k_T, m)
\,,\end{align}
and arises purely from its contribution to the gluon wave function renormalization;
here $\cJ_{g/g}^{(1,n_\ell)}(z, k_T)$ is the well-known NLO matching coefficient
for the gluon TMD onto the collinear gluon FF in the case
where both are defined in a purely light theory with $n_\ell$ flavors,
see \eq{matching_coeff_light_quark_limit}.
After $\MSbar$-renormalizing the matching coefficient, the net result is
\begin{align} \label{eq:cJ1gg_kT_ren_one_loop}
\Delta \cJ_{g/g}^{(1)}(z, k_T, m, \mu)
= \frac{\as T_F}{4\pi} \, \delta(1-z) \frac{1}{\pi}\delta(k_T^2) \, \frac{4}{3} \ln \frac{m^2}{\mu^2}
\,.\end{align}
This agrees with \refcite{Pietrulewicz:2023dxt},
where the same contribution was recently revisited
as part of computing all secondary heavy-quark effects
in the two-loop gluon TMD PDF;
we refer to that reference for a detailed discussion
of how it ensures the proper TMD renormalization properties
when connecting the theories with $n_\ell$ and $n_\ell + 1$ quark degrees of freedom.

Taking cumulant integrals, we find
\begin{align} \label{eq:nonvalence_kTcut_one_loop}
d_{1\,Q/g}^{(1)}(z, \ktcut)
&= \frac{\as}{4 \pi} T_F\frac{2}{  z^2} \biggl\{(2z^2-2z+1)\ln (\xcut z^2+1)+\frac{2 \xcut(1-z) z^3}{{\xcut} z^2+1}\biggr\}
\,, \nn \\
\cJ_{g/Q}^{(1)}(z, \ktcut, m)
 &= \frac{\as}{4 \pi} C_F  \frac{2}{z}\biggl\{(z^2-2 z+2)\ln (\xcut +1)+\frac{2 \xcut (z-1)}{1+\xcut}\biggr\}
\,,\end{align}
where $\xcut \equiv (\ktcut/m)^2$.
Finally, taking a $J_1$ Hankel transform of the cumulant results
as in \eq{hankel_transform_kTcut_to_bT},
we arrive at the following $b_T$-space results:
\begin{align} \label{eq:nonvalence_bT_one_loop}
d_{1\,Q/g}^{(1)}(z, b_T)
=
&    \frac{\as}{4 \pi} T_F\frac{4}{ z^2} \biggl\{(2z^2-2z+1) K_0\Bigl(\frac{b_T m}{z}\Bigr)+ b_T m (1-z) K_1\Bigl(\frac{b_T m}{z}\Bigr)\biggr\}
\,, \nn \\
\cJ_{g/Q}^{(1)}(z, b_T, m)
&=\frac{\as}{4 \pi} C_F  \frac{4}{z}\bigl\{
   (z^2-2 z+2) K_0(b_T m)
   - b_T m (1-z) K_1(b_T m)
\bigr\}
\,.\end{align}

\section{bHQET TMD fragmentation factors at NLO}
\label{sec:bhqet_tmd_fragmentation_factors}

\subsection{All-order renormalization properties}
\label{sec:bhqet_tmd_fragmentation_factors_renormalization}

In \refcite{vonKuk:2023jfd} we showed that up to a normalization factor $N_H$
involving Clebsch-Gordan coefficients,
the unpolarized bHQET TMD fragmentation factor $\chi_{1,H}$
in \eq{tmdffs_bhqet_all_order_final_results}
for a heavy hadron state $H$ is given
by the trace of the spin density matrix $\rho_{\ell}$
encoding the dynamics of the light constituents $\ell$ of the hadron,
\begin{align} \label{eq:chi1H_in_terms_of_rho_ell}
\chi_{1,H}(b_T, \mu, \rho)
= N_H \sum_{h_\ell}
\rho_{\ell, h_\ell h_\ell}(b_\perp, \mu, \rho)
\,.\end{align}
Here we use the shorthand $\rho \equiv \nbar \cdot v$
as introduced below \eq{chi1h_ope},
which can be interpreted as the total boost of the fragmenting heavy-quark system
with velocity $v$ in the frame
where the Wilson line direction $\nbar$ has unit spatial component.
A priori, the matrix element is invariant under boosts
-- or reparametrization invariance type-III transformations (RPI-III) in SCET~\cite{Manohar:2002fd, Marcantonini:2008qn} --
that would take $\nbar \to \alpha \, \nbar$ for an arbitrary factor $\alpha$,
but similarly to the case of relativistic TMD FFs,
the presence of rapidity divergences breaks this invariance
such that $\chi_{1,H}$ and $\rho_{\ell}$ develop an anomalous dependence on $\rho$.
In this subsection, we make the discussion of this point in \refcite{vonKuk:2023jfd}
fully explicit, allowing us to cross check our predictions
for the renormalization properties at one loop
in \sec{bhqet_tmd_fragmentation_factors_one_loop}.

Spelling out all regulators and its open light helicity indices $h_\ell, h_\ell'$,
the renormalized light-spin density matrix is defined as
\begin{align} \label{eq:def_rho_ell_ren}
\rho_{\ell, h_\ell h_\ell'}(b_\perp, \mu, \rho)
= \lim_{\eps \to 0} Z_{\chi_{1}}^{-1}(\mu, \rho, \eps)
\lim_{\eta \to 0} \Bigl[
\rho_{\ell, h_\ell h_\ell'}^\mathrm{bare}\Bigl(b_\perp, \eps, \eta, \frac{\rho}{\nu}\Bigr) \, \sqrt{S^{(n_\ell)}}(b_T, \eps, \eta, \nu)
\Bigr]
\,,\end{align}
where $S^{(n_\ell)}$ is the bare TMD soft function in the theory
where only the $n_\ell$ light quark degrees of freedom are active.
This is the same soft function as appears in light-quark TMD PDFs or FFs
at transverse momenta below the heavy-quark threshold:
The fact that it also arises here and cancels the rapidity divergences as $\eta \to 0$
in the bare light-spin density matrix
follows from consistency of the TMD factorization
for flavor-changing hard processes,
e.g.\ charged-current DIS $e^- \bar{s} \to \nu_e \, \bar{c}$.%
\footnote{
We assume as in \sec{theory_intro} that a rapidity regulator $\eta$
is used for which soft (zero-bin) subtractions
in the bare collinear matrix element are scaleless.
If this is not the case, it is crucial that zero-bin subtractions are defined
by modes that are soft in a frame where the heavy quark is highly boosted,
i.e., their interactions with the heavy-quark sector
are encoded in lightlike Wilson lines
and the multiplicative zero-bin subtraction factor
is obtained by taking $\nbar \cdot v \to \infty$.
}
The UV $\MSbar$ renormalization factor $Z_{\chi_{1}}$ likewise
follows from consistency and is given by
\begin{align} \label{eq:def_Zchi1}
Z_{\chi_{1}}^{-1}\bigl(\mu, \sqrt{\zeta}/m, \eps \bigr)
&= Z_\mathrm{UV}(\mu, \zeta, \eps) \, Z_{C_m}(m, \mu, \zeta, \eps)
\nn \\
&= 1 - \frac{\as C_F}{4\pi} \frac{1}{\eps} \Bigl(
   - 4 \ln \frac{\sqrt{\zeta}}{m} + 2
\Bigr)
+ \ord{\as^2}
\,,\end{align}
where $Z_\mathrm{UV}$ is the standard UV TMD renormalization factor
in \eqs{def_d1qq_bT_ren}{Z_uv_one_loop},
and $Z_{C_m}$ is the $\MSbar$ renormalization factor
of the Wilson coefficient in \eq{tmdffs_bhqet_all_order_final_results}
generated by integrating out the heavy quark~\cite{Hoang:2015vua},
\begin{align} \label{eq:def_ZCm}
C_m^\mathrm{bare}(m, \zeta, \eps)
&= Z_{C_m}(m, \mu, \zeta, \eps) \, C_m(m, \mu, \zeta)
\,.\end{align}
Both the soft function and the UV renormalization are independent
of the hadronic states, and thus universal for all possible traces of $\rho_\ell$,
i.e., they are the same for all possible TMD fragmentation factors $\chi_{1,H}$ (including polarized ones).
In particular, they are the same also
for the total TMD fragmentation factor $\chi_1$
that we consider in \sec{bhqet_tmd_fragmentation_factors_chi1}.
Note that while $Z_\mathrm{UV}$ and $Z_{C_m}$
individually depend on the dimensionful Collins-Soper scale $\zeta = (\nbar \cdot P_H/z)^2$,
consistency requires that only the dimensionless combination $\rho = \sqrt{\zeta}/m$
appears in $Z_{\chi_{1}}$ on the left-hand side of \eq{def_Zchi1}.
The fact that $\rho$ is dimensionless also implies
that $Z_{\chi_{1}}(\mu, \rho)$ can only subtract single poles in $\eps$,
but this is indeed precisely what one expects for the renormalization
of the product of Wilson lines with a timelike-lightlike cusp~\cite{Grozin:2015kna}.

Finally, the bare light-spin density matrix itself is defined as
\begin{align}
\rho_{\ell, h_\ell h_\ell'}^\mathrm{bare}\Bigl(b_\perp, \eps, \eta, \frac{\nbar \cdot v}{\nu}\Bigr)
&= \frac{1}{N_c} \Tr
\SumInt_{X}
\Mae{0}{W_{\eta}^\dagger(b_\perp) \,
   Y_v(b_\perp) \Ket{s_\ell, h_\ell, f_\ell ; X}
\nn \\ & \qquad \qquad \quad
   \times \Bra{s_\ell, h_\ell', f_\ell ; X} Y^\dagger_v(0) \,
W_{\eta}(0)}{0}
\,,\end{align}
where $\ket{s_\ell, h_\ell, f_\ell}$ indicates a light constituent state $\ell$
of total spin $s_\ell$, spin component $h_\ell$ along the direction
$z^\mu \equiv v^\mu - \frac{\bn^\mu}{\bn \cdot v}$,
and valence flavor content $f_\ell$.
The Wilson lines $Y_v$ along the timelike direction
encode the static interactions with the heavy quark,
\begin{align}
Y_v(x) \equiv P \Bigl[ \exp \Bigl(
   \img g \int_0^\infty \! \df s \, v \cdot A(x + v s)
\Bigr) \Bigr]
\,.\end{align}
The rapidity regulator only acts on the Wilson lines $W_\eta$
along the lightlike direction $\nbar$, which are still defined as in \eq{def_wilson_line}.
Here the rapidity regulator is the only source of boost invariance (RPI-III) breaking,
which implies that the bare spin density matrix can only depend on the combination
$\rho/\nu = \nbar \cdot v/\nu$.
A priori, it is not obvious that $\rho = \sqrt{\zeta}/m$,
as it appears in the renormalization factor in \eq{def_Zchi1},
should coincide with the $\rho = \nbar \cdot v$ that appears in the bare matrix element,
but this is in fact guaranteed by the crucial relation
\begin{align}
\sqrt{\zeta} = \nbar \cdot P_H /z = m \, \nbar \cdot v
\end{align}
that arises from consistently expanding to leading power in $1/m$~\cite{vonKuk:2023jfd}.

We conclude this section by summarizing the renormalization group equations
for the renormalized $\rho_\ell$ in $d = 4$ dimensions
that follow from the above renormalization procedure:
\begin{align} \label{eq:rho_ell_rge}
\mu \frac{\df}{\df \mu} \rho_{\ell, h_\ell h_\ell'}(b_\perp, \mu, \rho)
&= \gamma_{\chi_1}\bigl[ \as(\mu), \rho \bigr] \, \rho_{\ell, h_\ell h_\ell'}(b_\perp, \mu, \rho)
\,, \nn \\
\rho \frac{\df}{\df \rho} \rho_{\ell, h_\ell h_\ell'}(b_\perp, \mu, \rho)
&= \gamma_\zeta^{(n_\ell)}(b_T, \mu) \, \rho_{\ell, h_\ell h_\ell'}(b_\perp, \mu, \rho)
\,,\end{align}
where the $\mu$ anomalous dimension is given by the difference
of the standard TMD UV anomalous dimension $\gamma_\mu^q$
and the anomalous dimension of $C_m$,
evaluating both at $\zeta = \rho^2 m^2$,
\begin{align}
\gamma_{\chi_1}\bigl[ \as(\mu), \rho \bigr]
= \gamma_\mu^q(\mu, \rho^2 m^2) - \gamma_{C_m}(\mu, m^2, \rho^2 m^2)
= \frac{\as C_F}{4\pi} \Bigl( - 8\ln \rho + 4 \Bigr) + \ord{\as^2}
\,.\end{align}
We stress that when extracting the anomalous dimension at higher orders,
care must be taken to perturbatively reexpress $\as^{(n_\ell + 1)}$
in terms of $\as(\mu) \equiv \as^{(n_\ell)}(\mu)$.

From the second line of \eq{rho_ell_rge}, we see that $\rho_\ell$ develops
an anomalous dependence on the boost parameter $\rho$ governed
by the Collins-Soper CS kernel $\gamma_\zeta^{(n_\ell)}$
of the theory where only $n_\ell$ light quarks are active.
For perturbative $b_T$, it is given in our normalization by
\begin{align}
\gamma_\zeta^{(n_\ell)}(b_T, \mu)
= \frac{\as C_F}{4\pi} \, \biggl(
   - 8 \ln \frac{b_T \mu}{2 e^{-\gamma_E}}
\biggr) + \ord{\as^2} + \ord{\lqcd^2 b_T^2}
\,.\end{align}
A natural boundary condition for $\rho_\ell$, where it is free of large rapidity logarithms,
is given by $\rho \sim 1$, i.e., the rest frame of the heavy quark.
By contrast, in physical cross sections, the boost is fixed to $\rho = Q/m \gg 1$,
i.e., the lab frame in the $e^+ e^-$ case, or the Breit frame for SIDIS.
The $\rho$ evolution between those scales resums
the associated large rapidity logarithms.
As in the relativistic case, the two-dimensional evolution equations satisfy
an integrability condition,
\begin{align}
\rho \frac{\df}{\rho} \gamma_{\chi_1}\bigl[ \as(\mu), \rho \bigr]
= \mu \frac{\df}{\mu} \gamma_\zeta^{(n_\ell)}(\mu, b_T)
= -2 \Gamma_\mathrm{cusp}\bigl[ \as(\mu) \bigr]
\end{align}
which involves the lightlike-lightlike cusp anomalous dimension
$\Gamma_\mathrm{cusp}(\as) = \tfrac{\as}{4\pi} \, 4 C_F + \ord{\as^2}$
and can readily be verified at one loop.

\subsection{Sum over hadronic states}
\label{sec:bhqet_tmd_fragmentation_factors_chi1}

An important property of the unpolarized bHQET fragmentation factors
is that their description greatly simplifies
when summing over all hadrons $H$ containing the heavy quark,
\begin{align}
\chi_1(b_T, \mu, \rho)
= \sum_H \chi_{1,H}(b_T, \mu, \rho)
= \sum_\ell \sum_{h_\ell} \rho_\ell(b_\perp, \mu, \rho)
\,.\end{align}
As indicated, this sum in turn is equal to the sum of traces of the light-spin density matrices
for all the possible light states $\ell$.
Unlike the case of inclusive fragmentation, $\chi_1$ does not simply
evaluate to unity by probability conservation.
However, \emph{as} in the inclusive case, the complete sum over states
can be performed also here, leaving behind a simple vacuum
Wilson line correlator~\cite{vonKuk:2023jfd},
\begin{align} \label{eq:def_chi1_bare}
\chi_1^\bare(b_T,\eps, \eta, \nbar\cdot v)
\equiv
\frac{1}{N_c} \Tr
\Mae{0}{W_{\eta}^\dagger(b_\perp) \,
   Y_v(b_\perp) \, Y^\dagger_v(0) \,
W_{\eta}(0)}{0}
\,.\end{align}
Its renormalization proceeds exactly as for the whole $\rho_\ell$ in \eq{def_rho_ell_ren},
\begin{align} \label{eq:def_chi1_ren}
\chi_1(b_T, \mu, \rho)
= \lim_{\eps \to 0} Z_{\chi_{1}}^{-1}(\mu, \rho, \eps)
\lim_{\eta \to 0} \Bigl[
\chi_1^\mathrm{bare}\Bigl(b_T, \eps, \eta, \frac{\rho}{\nu}\Bigr) \, \sqrt{S^{(n_\ell)}}(b_T, \eps, \eta, \nu)
\Bigr]
\,.\end{align}

While \eqs{def_chi1_bare}{def_chi1_ren} are valid nonperturbatively
and likely exhibit interesting properties at long distances $b_T \sim 1/\lqcd$,
an immediate application of their simple operator structure is
to evaluate the correlator perturbatively at $b_T \ll 1/\lqcd$,
which we do in the next section.
As a byproduct of this, we may then read off the universal
matching coefficient $C_1$ in \eq{chi1h_ope}
by summing \eq{chi1h_ope} over all hadrons $H$
such that $\sum \chi_H = 1$ drops out,
\begin{align}
\chi_1(b_T, \mu, \rho) = C_1(b_T, \mu, \rho) + \ord{\lqcd^2 b_T^2}
\,.\end{align}
Of course, one can also determine $C_1$ by a standard matching calculation,
e.g.\ using non-decoupled heavy-quark fields $h_v$ in the operators
and heavy-quark external states in both the theory at the scale $\mu \sim 1/b_T$
and at the low scale $\mu \sim \lqcd$ to read off the matching coefficient,
and the procedure (in pure dimensional regularization) involves equivalent diagrams
and leads to the same result.

\subsection{One-loop calculation}
\label{sec:bhqet_tmd_fragmentation_factors_one_loop}

\begin{figure*}
\centering
\subfloat[]{
   \includegraphics[width=0.22\textwidth]{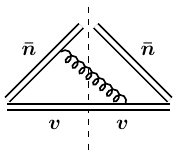}
}%
\subfloat[]{
   \includegraphics[width=0.22\textwidth]{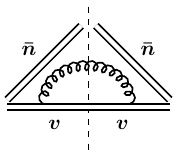}
}%
\raisebox{-4em}{
   \includegraphics[width=0.48\textwidth]{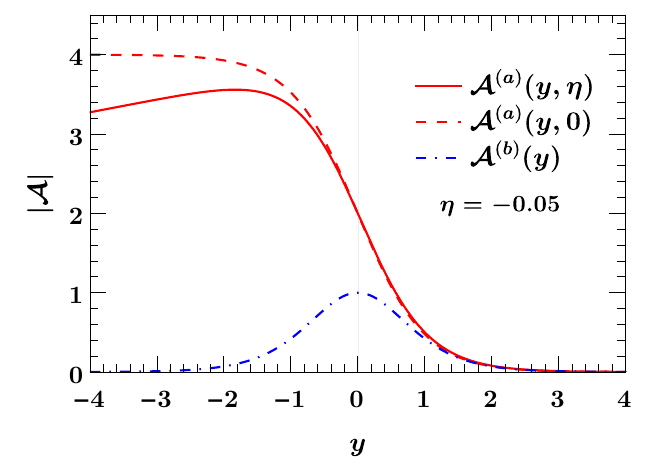}
}
\caption{
Left: Diagrams contributing to the total bHQET TMD fragmentation factor
in the perturbative domain.
Wilson lines along the lightlike $\bn$ direction are indicated
by the slanted double lines.
The horizontal double lines indicate
the Wilson lines along the timelike $v$ direction.
The dashed line indicates on-shell cuts
after introducing a complete set of states.
The mirror diagram for (a) is understood
and included in expressions given in the text.
Right: Phase-space integrands $\cA$ for the two real-emissions diagrams
as a function of the gluon emission rapidity $y$.
}
\label{fig:bhqet_diagrams}
\end{figure*}

To evaluate $\chi_1$ perturbatively, it is convenient
to reinstate a complete sum over states and transition to momentum space
to evaluate the bare matrix element,
\begin{align} \label{eq:def_chi1_kT_bare}
\chi_1^\bare\Bigl(k_T, \eps, \eta, \frac{\nbar\cdot v}{\nu}\Bigr)
=
\frac{1}{N_c} \Tr
\int \! \frac{\df \Omega_{2-2\eps}}{2\pi} \, k_T^{-2\eps}
\Mae{0}{
   W_{\eta}^\dagger \, Y_v
}{X}
\Mae{X}{\bigl[
   \delta^{(2-2\eps)}\bigl(k_\perp^\mu + \img\partial_\perp^\mu\bigr) \,
   Y^\dagger_v \, W_{\eta}
\bigr]}{0}
\,,\end{align}
in analogy to \eq{def_d1qq_kT_bare}.
Virtual diagrams with $X$ the vacuum state are scaleless.
The two contributing real emission diagrams at $\ord{\as}$
in Feynman gauge are shown in \fig{bhqet_diagrams}.
The one-gluon Feynman rule for the $Y_v$ Wilson line
can e.g.\ be read off from \refcite{Fleming:2007xt},
which uses a covariant derivative convention
consistent with the one-gluon Feynman rule we use for $W_\eta$.
Including the mirror diagram for diagram~(a),
the two diagrams are both given by simple integrals
over the rapidity $y$ of the emitted gluon
in the heavy-quark rest frame,
\begin{align}
\chi_1^{(a)}
&= \frac{\as C_F}{4\pi} \frac{e^{\eps \gamma_E}}{\Gamma(1-\eps)}
\frac{\mu^{2\eps}}{\pi k_T^{2 + 2\eps + \eta}} \Bigl( \frac{\nbar \cdot v}{\nu} \Bigr)^{-\eta}
\int \! \df y \, \cA^{(a)}(y, \eta)
\,, \nn \\
\chi_1^{(b)}
&= \frac{\as C_F}{4\pi} \frac{e^{\eps \gamma_E}}{\Gamma(1-\eps)}
\frac{\mu^{2\eps}}{\pi k_T^{2 + 2\eps}} \int \! \df y \, \cA^{(b)}(y)
\,,\end{align}
where the relevant integrands are given by
\begin{align}
\cA^{(a)}(y, \eta)
= \frac{2 e^{-\eta y}}{e^y \, \cosh y}
\,, \qquad
\cA^{(b)}(y)
= - \frac{1}{\raisebox{-0.1em}{$\cosh^2 y$}}
\,,\end{align}
and their integrals evaluate to
\begin{align}
\int \! \df y \, \cA^{(a)}(y, \eta)
= - \frac{2\pi}{\sin(\eta \pi/2)} = - \frac{4}{\eta} + \ord{\eta}
\,, \qquad
\int \! \df y \, \cA^{(b)}(y)
= -2
\,.\end{align}
The origin of the rapidity divergence, like the structure of the integrands
as a whole, has a rather intuitive interpretation, which we illustrate
in the right panel of \fig{bhqet_diagrams}:
Emissions at large positive gluon rapidity $y$
in the rest frame of the heavy quark are strongly suppressed
for either contributing diagram,
which is the well-known dead-cone effect~\cite{Dokshitzer:1991fd}.
For diagram~(b), where the static heavy quark $Y_v$ both emits and absorbs the gluon,
the contribution to the emission probability (dot-dashed blue) peaks at $y = 0$
and is suppressed also at large negative rapidities
because the contribution must be even in the rest frame.
In contrast to this, the contribution from diagram~(a),
where the emission couples to the $W_\eta$
that arises from integrating out the interactions
with the anticollinear and soft sectors
and is required for gauge invariance,
tends to a large constant as $y \to -\infty$ (dashed red),
leading to a rapidity divergence.
Of course, this precisely corresponds to the central soft momentum region,
and thus after introducing the regulator $e^{-\eta y}$ with $\eta < 0$
to dampen the behavior as $y \to \infty$ (solid red),
the associated pole cancels against the soft function as usual.

Adding the one-loop soft function in \eq{bare_tmd_soft_function},
which at this order is independent of the matter content of the theory,
we find that the $\eta$ poles cancel for arbitrary $\eps$ as they must.
Taking $\eta \to 0$, we find, using the tree-level result
$\chi_1^{(0)}(k_T) = \tfrac{1}{\pi} \delta(k_T^2)$,
\begin{align}
&\lim_{\eta \to 0} \Bigl[
   \chi_1^{\bare\,(1)}(k_T, \eps, \eta, \frac{\nbar\cdot v}{\nu}\Bigr)
   + \tfrac{1}{2} S^{(n_\ell)\,(1)}(k_T, \eps, \eta, \nu)
\Bigr]
\nn \\
&= \frac{\as C_F}{4\pi} \frac{e^{\eps \gamma_E}}{\Gamma(1-\eps)}
\frac{\mu^{2\eps}}{\pi k_T^{2 + 2\eps}}\Bigl(4 \ln \nbar \cdot v - 2 \Bigr)
\nn \\
&= \frac{\as C_F}{4\pi} \, \frac{1}{\pi} \Bigl[
   - \frac{\delta(k_T^2)}{\eps} + \cL_0(k_T^2, \mu) + \ord{\eps}
\Bigr]
\Bigl(4 \ln \nbar \cdot v - 2 \Bigr)
\,.\end{align}
The remaining pole in $\eps$ is precisely canceled by the predicted counter term in \eq{def_Zchi1},
confirming the renormalization structure we discussed in \sec{bhqet_tmd_fragmentation_factors_renormalization} at one loop.
Taking a Fourier transform of the finite remainder,
our final result for the renormalized total TMD fragmentation factor
in the perturbative domain, and thus for the matching coefficient $C_1$, reads
\begin{align} \label{eq:chi1_C1_final_result}
\chi_1(b_T, \mu, \rho)
&= C_1(b_T, \mu, \rho) + \ord{\lqcd^2 b_T^2}
\nn \\
&= 1 + \frac{\as C_F}{4\pi} (-L_b) \bigl( 4\ln \rho - 2\bigr)
+ \ord{\as^2} + \ord{\lqcd^2 b_T^2}
\,,\end{align}
where $L_b$ was defined in \eq{def_Lb_y_Ly}.

\section{Consistency checks in the large and small-mass limits}
\label{sec:consistency_relations}

\subsection{Large-mass limit}
\label{sec:heavy_quark_limit_consistency}

To check the behavior of our final position-space results for $d_{1\,Q/Q}$
in \eq{d1QQ_bT},
we first note that the result is exponentially suppressed at $z < 1$ for $y \equiv m b_T \gg 1$
as a consequence of \eq{how_we_count_zH}
and the large-argument behavior of the Bessel $K$ functions,
i.e., it is given by $\delta(1-z)$ up to a proportionality factor.
To verify the prediction from the heavy-quark limit in \eq{tmd_ff_unpol_consistency_heavy},
it is then sufficient to compare the total $z$ integral
(or any other $z$ moment) of both sides of \eq{tmd_ff_unpol_consistency_heavy}.
The total $z$ integral yields
\begin{align} \label{eq:d1QQ_bT_large_mass}
\int \! \df z \, d_{1\,Q/Q}^{(1)}(z, b_T, \mu, \zeta)
&= \frac{\as C_F}{4\pi} \biggl\{
- 2 L_b \ln\frac{\zeta}{m^2}
+ 4 \ln^{2}\frac{\mu}{m}
+ 6 \ln \frac{\mu}{m}
+ 4 - L_y^2 - \frac{\pi^2}{6}
\nn \\
& \qquad \qquad
- \frac{4\pi}{y} + 4 \cL_0(y^2, b_0^2) + 12 \delta(y^2) + \ORd{\frac{1}{y^3}}
\biggr\}
\,.\end{align}
Here we used \eq{cR_total_z_integral} for the total $z$ integral
of the cumulant-space remainder term $\cR(z, \ktcut/m)$,
which we then power expanded for small $\ktcut/m$,
letting us evaluate the total $z$ integral
of \eq{def_tcR} power by power in $1/m$.
We find that the leading $\ord{1/m^0}$ term on the first line is in full agreement
with the bHQET prediction from \eq{tmd_ff_unpol_consistency_heavy},
using our one-loop result for $C_1$ in \eq{chi1_C1_final_result}
and the known result for $C_m$ in \eq{wilson_coeff_Cm}.
We point out the interesting observation
that the first subleading $\ord{1/m}$ correction in \eq{d1QQ_bT_large_mass}
is free of large logarithms $L_y \equiv 2\ln \tfrac{y}{b_0}$, $b_0 \equiv 2 e^{-\gamma_E}$,
suggesting a simple structure if factorized in terms
of subleading $\ord{1/m}$ TMD matrix elements in bHQET.
On the other hand, the terms at the next order
require plus regularization in $b_T$ space
due to their $1/y^2$ scaling
and do feature a single-logarithmic term $\cL_0(y^2, b_0^2)$.

For completeness, we may also expand the nonvalence results in \eq{nonvalence_bT_one_loop}
for $y \gg 1$ at finite values of $0 < z < 1$ in a similar way,
which simply yields
\begin{align}
d_{1\,Q/g}^{(1)}(z, b_T)
&= \frac{\as T_F}{4 \pi} \frac{1}{z^2} \biggl\{
   8 z^2 \,\delta(y^2) + \ORd{\frac{1}{y^3}}
\biggr\}
\,, \nn \\
\cJ_{g/Q}^{(1)}(z, b_T, m)
&=\frac{\as C_F}{4 \pi} \biggl\{
   8 z \, \delta(y^2) + \ORd{\frac{1}{y^3}}
\biggr\}
\,,\end{align}
i.e., both channels are in fact suppressed by $1/m^2$ at finite $z$.

\subsection{Small-mass limit}
\label{sec:light_quark_limit_consistency}

Expanding \eqs{d1QQ_bT}{nonvalence_bT_one_loop}
for small values of the mass, $y \equiv m b_T \ll 1$, we obtain:
\begin{align}
d^{(1)}_{1\,Q/Q}(z_H, b_T, \mu, \zeta)
&= \frac{\as C_F}{4\pi} \frac{1}{z^2}
\bigg\{  \delta(1-z) \bigg [
- 2 L_b \ln\frac{\zeta}{m^2}
+ 4 \ln^{2}\frac{\mu}{m}
+ 6 \ln \frac{\mu}{m}
- L_y^2
+ 4 - \frac{\pi^{2} }{6}
\bigg]
\nn \\[0.2em]
& \qquad \qquad \qquad
- 4 (1 + L_y ) \, \cL_{0}(1-z)
- 8 \cL_{1}(1-z)
\nn \\
& \qquad \qquad \qquad
+ 4 + 2 L_y(1 + z) + \frac{8 \ln \zb - 4(1 + z^2)\ln \tfrac{\zb}{z}}{\zb}
\nn \\
& \qquad \qquad \qquad
+ y^2 \biggl(
   \frac{1}{z^2} - z - \frac{L_y(1+z)^2 \zb}{2z^2}
   - \frac{(1 + z)^2 \zb \ln \tfrac{\zb}{z}}{z^2} \biggr)
   +\ord{y^4}
\bigg\}
\,, \nn \\
d^{(1)}_{1\,Q/g}(z_H, b_T, \mu, \zeta)
&= \frac{\as T_F}{4\pi} \biggl\{
   -2 L_y [2 (z-1) z+1]-4 (z-1) z + 4 [2 (z-1) z+1] \ln z
\nn \\
& \qquad \qquad \qquad
   + y^2 \biggl(-\frac{L_y (1-2 z)^2}{2 z^2}+\frac{3 (z-1) z+1}{z^2} + \frac{(1-2 z)^2 \ln z}{z^2}\biggr)
   + \ord{y^4}
\biggr\}
\,, \nn \\
\cJ^{(1)}_{g/Q}(z_H, b_T, \mu, \zeta)
&= \frac{\as C_F}{4\pi} \biggl\{
   -\frac{2 L_y [(z-2) z+2]}{z}+\frac{4 (z-1)}{z}
\nn \\
& \qquad \qquad \qquad
   + y^2 \left(-\frac{L_y (z-2)^2}{2 z}+z+\frac{3}{z}-3\right)
   + \ord{y^4}
\biggr\}
\,.\end{align}
It is straightforward to verify that the leading $\ord{y^0}$ terms
are in full agreement with the prediction in \eq{consistency_relations_small_mass}
using the one-loop ingredients collected in our notation
in \App{perturbative_ingredients}.
At leading power, the logarithms $L_y \equiv 2 \ln \tfrac{y}{2e^{-\gamma_E}}$
in all channels are predicted by the timelike DGLAP evolution
between the TMD matching coefficients at $\mu \sim 1/b_T$ and the matching kernels
and partonic collinear FFs at $\mu \sim m$.
The $\ord{y^2}$ terms, which we here give explicitly,
likewise feature single logarithms $L_y$ in all partonic channels,
and can be used to investigate the $\ord{b_T^2}$ twist-4 matching of TMD FFs
in a physics setup where the low scale matrix elements at $\mu \sim m$
are perturbatively calculable.

\section{Joint threshold and \texorpdfstring{$q_T$}{qT} resummation for massive quark fragmentation}
\label{sec:joint_res}

\begin{figure*}
\centering
\subfloat[]{
   \includegraphics[width=\WidthThreeSubfigs]{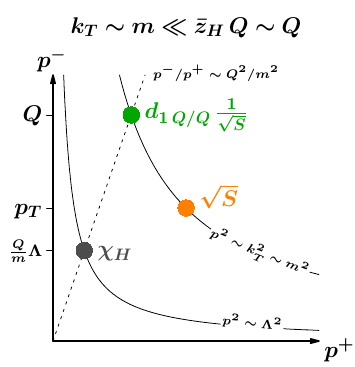}
}%
\subfloat[]{
   \includegraphics[width=\WidthThreeSubfigs]{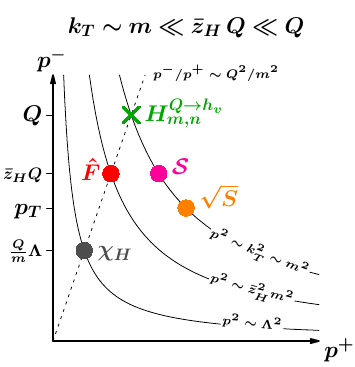}
}%
\subfloat[]{
   \includegraphics[width=\WidthThreeSubfigs]{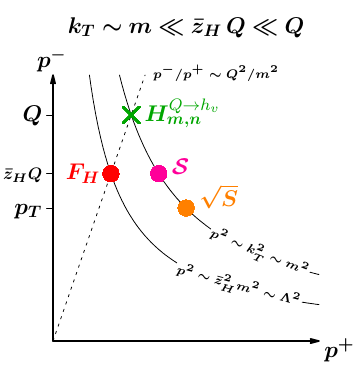}
}
\caption{
Relevant modes contributing to the heavy-quark TMD FF
as the collinear momentum fraction $z_H \to 1$
while counting $m \sim k_T$.
See the text for a detailed discussion.
}
\label{fig:modes_joint_resummation}
\end{figure*}

The remarkable simplicity of our final $b_T$-space result in \eq{d1QQ_bT}
as $z \to 1$ warrants further investigation.
Specifically, the leading terms at $\ordsq{(1-z)^{-1}}$ feature
only a logarithmic dependence on the mass and $b_T$,
while in general, a power-like dependence on $m b_T \sim 1$ would be allowed,
suggesting that the heavy-quark TMD FF can be factorized further
in this regime such that the logarithms become predicted by RGEs.
To investigate this question, we modify our previous assumption
on the power counting of $\zb_H \equiv 1 - z_H$, \eq{how_we_count_zH},
and instead count
\begin{align} \label{eq:how_we_count_zH_for_joint_res}
\text{\Sec{joint_res}:} \quad \frac{k_T}{Q} \sim \frac{m}{Q} \ll \zb_H \ll 1
\,.\end{align}
We stress that we \emph{continue} to count $m \sim k_T \sim 1/b_T$.
In \fig{modes_joint_resummation} we illustrate the contributing modes
in the EFT for this scenario.
\Fig{modes_joint_resummation}~(a) shows the original, unmodified mode setup
that led to \eq{tmd_ff_unpol_bhqet}.
Here the partonic heavy-quark TMD FF arises from matching collinear modes (green),
accounting for the usual contribution of the central soft modes (orange),
onto the nonperturbative modes (gray) describing the formation of the brown muck,
which is isotropic in the rest frame of the heavy quark indicated by the dashed black line.
The collinear modes are kinematically unconstrained
in the longitudinal direction because $\zb_H \sim 1$.

Once the modified assumption \eq{how_we_count_zH_for_joint_res} is made,
energetic collinear modes at the scale $\mu \sim k_T \sim m$
can no longer be radiated into the final state,
see \fig{modes_joint_resummation}~(b) and (c).
Their effect becomes purely virtual and is encoded
in a matching coefficient $H_{m,n}^{Q \to h_v}$ (green cross)
from matching onto bHQET modes at lower virtuality.
Real emissions into the $n$-collinear direction at the scale $\mu \sim k_T \sim m$
can instead only have collinear-soft (pink) scaling
\begin{align}
p_\mathrm{cs} \sim \Bigl(\zb_H Q, \tfrac{k_T^2}{\zb_H Q}, k_T\Bigr)
\,,\end{align}
which is the characteristic mode of double-differential $z \to 1$ threshold
and $q_T$ resummation, often simply dubbed
joint resummation~\cite{Li:1998is, Laenen:2000ij, Kulesza:2002rh, Kulesza:2003wn, Banfi:2004xa, Bozzi:2007tea, Debove:2011xj, Li:2016axz, Lustermans:2016nvk}.%
\footnote{
We note that while \refcite{Banfi:2004xa}
also dealt with final-state heavy quarks,
that work considered the joint resummation for the total transverse momentum
of a heavy quark \emph{pair} with respect to the beam direction,
which is a distinct problem from the one we consider here.
}

The physical picture at scales $\mu \ll k_T \sim m$ in this scenario,
which is governed by one or several bHQET modes
at their characteristic lab-frame rapidity $p^-/p^+ \sim Q^2/m^2$,
further depends on the relative size of $\lqcd/m$ and $\zb_H$,
specifically whether $\lqcd/m \sim \zb_H$ or $\lqcd/m \ll \zb_H$.
(Since $\lqcd$ introduces a fourth scale into the problem,
i.e., a third independent dimensionless power counting parameter,
either case may hold in general and independent of $k_T/Q$ and $m/Q$.)
The relevant modes for the case $\lqcd/m \sim \zb_H$
are illustrated in \fig{modes_joint_resummation}~(c),
and are in fact analogous to the well-understood case
of single-inclusive heavy-hadron production near the endpoint in $e^+e^-$ collisions~\cite{Neubert:2007je, Fickinger:2016rfd}.
Specifically, the nonperturbative fragmentation process
at the scale $\mu \sim \lqcd \sim \zb_H m$
provides a leading contribution to the measured longitudinal
hadron momentum in this case (but not to its transverse momentum,
where the contribution is power suppressed!).
This effect is encoded in the well-known collinear bHQET fragmentation
shape function $F_H(k^-/v^-, \mu)$~\cite{Neubert:2007je, Fickinger:2016rfd}.%
\footnote{We use the symbol $F_H$ here to distinguish the shape function
from the TMD soft function; in \refcite{Fickinger:2016rfd},
the symbol $S_{H/Q}$ is used instead,
and our partonic $\hat{F}$ corresponds to their $S_{Q/Q}$.}
In total, the factorization of the hadron-level heavy-quark TMD FF
in this regime reads, with $\zeta = \omega^2$ the Collins-Soper scale,
\begin{align} \label{eq:joint_res_heavy_shape_function_np}
D_{1\,H/Q}(z_H, b_T, \mu, \omega^2)
&= H_{m,n}^{Q \to h_v}\Bigl(m, \mu, \frac{\nu}{\omega} \Bigr) \, \int \! \df \ell^- \, \cS(\ell^-, b_T, m, \mu, \nu ) \, \sqrt{S}(b_T, m, \mu, \nu)
\nn \\ & \quad
\times \int \! \frac{\df k^-}{v^-} \, F_H\Bigl(\frac{k^-}{v^-}, \mu\Bigr) \, \delta\Bigl(\zb_H - \frac{ \ell^- + k^-}{\omega} \Bigr)
\nn \\ & \quad
\times \biggl[ 1 + \ord{\zb_H} + \ORd{\frac{\lqcd}{k_T}} + \ORd{\frac{\lqcd}{m}} \biggr]
\,,\end{align}
where $\cS$ and $\sqrt{S}$ are the separately rapidity-renormalized
collinear-soft function from joint resummation~\cite{Lustermans:2016nvk}
and the square root of the standard TMD soft function, respectively.
Similarly, $H_{m,n}^{Q \to h_v}$ (defined in \refcite{Hoang:2015vua} without the superscript,
which we here include for clarity)
is the rapidity-renormalized matching coefficient encoding
the virtual contributions of massive collinear quark modes $Q$
with label momentum $\omega$ matched onto bHQET fields $h_v$.
As in \refcite{Hoang:2015vua}, the effect of these modes is purely virtual
due to the kinematic constraint $z_H \to 1$.
We note that $H_{m,n}^{Q \to h_v}$ also includes the effect of secondary
heavy quark loops not directly connected to the operator starting at $\ord{\as^2}$,
which are responsible for its rapidity renormalization.
By evolving all terms in \eq{joint_res_heavy_shape_function_np}
from their canonical scales (e.g.\ in Mellin space $N \leftrightarrow z_H$,
where $\nu_\cS \sim \mu_{F_H} v^- \sim \omega/N$
and the evolution is multiplicative) to the common scale $\mu \sim \nu \sim Q$
with the hard function, all large threshold and transverse momentum
logarithms present in the heavy-hadron TMD cross section
in this regime can be resummed simultaneously.

It is instructive to compare the above to the refactorization
of the light-parton TMD FF in the joint resummation regime $k_T \ll \zb_h Q \ll Q$,
which follows from crossing
the light-parton TMD PDF refactorization relations \cite{Lustermans:2016nvk},
accounting for the additional effect of secondary heavy quarks starting at two loops,
\begin{align} \label{eq:joint_res_light}
D_{1\,h/q}(z_h, b_T, \mu, \omega^2)
&= H_{m,n}^{q \to q}\Bigl( m, \mu, \frac{\nu}{\omega} \Bigr) \int \! \df \ell^- \, \cS(\ell^-, b_T, m, \mu, \nu ) \, \sqrt{S}(b_T, m, \mu, \nu)
\nn \\ & \quad
\times \int \! \df k^- \, D^\mathrm{thr}_{h/q}\Bigl(\frac{k^-}{P_h^-}, \mu\Bigr) \, \delta\Bigl(\zb_h - \frac{ \ell^- + k^-}{\omega} \Bigr)
\nn \\ & \quad
\times \biggl[ 1 + \ord{\zb_H} + \ORd{\frac{\lqcd}{k_T}} + \ORd{\frac{\lqcd}{m}} \biggr]
\,,\end{align}
where $D^\mathrm{thr}_{h/q}$ is the collinear (threshold) fragmentation function,
whose renormalization is given by the timelike DGLAP kernels expanded to leading power in $1-z$.
The matching coefficient $H_{m,n}^{q \to q}$
(defined as $H_c$ in \refcite{Pietrulewicz:2017gxc})
encodes the virtual contributions of secondary heavy-quark loops
to the matching of light-quark collinear modes $q$ at the scale $\mu \sim m$
onto their respective counterparts in the theory at lower scales.
By comparing \eqs{joint_res_heavy_shape_function_np}{joint_res_light}
and making use of known results
for the heavy-quark threshold FF~\cite{Fickinger:2016rfd},%
\footnote{
Specifically, \refcite{Fickinger:2016rfd}
features a combined mass-mode matching coefficient $C_m^\text{ref.\,\cite{Fickinger:2016rfd}}(m, \mu)$
appearing as $D_{H/Q}(z_H = 1 - k^-/P_H^-, \mu)
= C_m^\text{ref.\,\cite{Fickinger:2016rfd}}(m, \mu) \, F_H(k^-/v^-, \mu)
\bigl[ 1 + \ord{\zb_H} \bigr]$,
not to be confused with our $C_m(m, \mu, \zeta)$ in \eq{tmdffs_bhqet_all_order_final_results}.
This relation, which defines $C_m^\text{ref.\,\cite{Fickinger:2016rfd}}$
as the finite remainder of the collinear FF after dividing out the shape function,
was then used to extract it to two loops
from the partonic heavy-quark FF~\cite{Melnikov:2004bm}.
We have verified to two loops
that $X \equiv C_m^\text{ref.\,\cite{Fickinger:2016rfd}}(m, \mu)/ [
H_{m,n}^{Q\to h_v}(m, \mu, \omega/\nu) / H_{m,n}^{q \to q}(m, \mu, \omega/\nu)
]$
is RG invariant, which together with \eq{joint_res_light}
guarantees RG invariance of \eq{joint_res_heavy_shape_function_np}.
In particular, the $\nu$ dependence cancels between $H_{m,n}^{Q\to h_v}$
and $H_{m,n}^{q \to q}$, as expected because it arises
from the common, secondary mass effects.
We stress, however, that the defining relation for $ C_m^\text{ref.\,\cite{Fickinger:2016rfd}}$
cannot simply be inserted back into \eq{joint_res_light}
because it only holds when all other scales in the problem are large compared to the mass.
Accordingly, $X$ does not need to be equal to one,
and we indeed find a nontrivial constant term at $\ord{\as^2}$.
}
we can conclude that \eq{joint_res_heavy_shape_function_np}
as the main new result of this section
is indeed RGE consistent.
We note that both \eqs{joint_res_heavy_shape_function_np}{joint_res_light}
also generically feature a dependence on the heavy quark mass
in the soft and collinear-soft function
since $m b_T \sim 1$, to which we return towards the end of this section.

To make contact with the \emph{partonic} heavy-quark TMD FF,
we next take $\lqcd \to 0$, which in particular means
that now $\lqcd/m \ll \zb_H$.
The mode setup for this regime is illustrated in \fig{modes_joint_resummation}~(b).
The only change compared to \fig{modes_joint_resummation}~(c)
is that the fragmentation shape function now factorizes
into the perturbative bHQET shape function $\hat{F}$
and the total fragmentation probability $\chi_H$~\cite{Fickinger:2016rfd},
\begin{align} \label{eq:ope_fragmentation_shape_function}
F_H\Bigl(\frac{k^-}{v^-}, \mu\Bigr) = \hat{F}\Bigl(\frac{k^-}{v^-}, \mu \Bigr) \, \chi_H + \ord{\lqcd}
\,.\end{align}
This implies that the joint resummation of the partonic heavy-quark TMD FF
is governed by the following refactorization relation,
\begin{align} \label{eq:joint_res_heavy_partonic}
d_{1\,Q/Q}(z, b_T, \mu, \omega^2)
&= H_{m,n}^{Q \to h_v}\Bigl( m, \mu, \frac{\nu}{\omega} \Bigr) \, \int \! \df \ell^- \, \cS(\ell^-, b_T, m, \mu, \nu ) \, \sqrt{S}(b_T, m, \mu, \nu)
\nn \\ & \quad
\times \int \! \frac{\df k^-}{v^-} \, \hat{F}\Bigl( \frac{k^-}{v^-}, \mu \Bigr) \, \delta\Bigl(\zb - \frac{ \ell^- + k^-}{\omega} \Bigr) + \Ordsq{(1-z)^0}
\,.\end{align}
This is precisely the relation governing the leading $1/\zb$ terms
in the partonic heavy-quark TMD FF that we were seeking out,
and it is straightforward to verify \eq{joint_res_heavy_partonic}
at one-loop order against \eq{d1QQ_bT}
using the fact that $\tcR(z, y) = \ordsq{(1-z)^0}$,
the fact that $v^- = \w/m$,
and the ingredients given in \app{perturbative_ingredients}.
In particular, since the soft and collinear-soft function
are defined as Wilson line correlators, they can only acquire mass dependence
through secondary mass effects starting at $\ord{\as^2 C_F T_F}$.
In turn, this implies that the mass dependence at $\ord{\as}$
arises only from that of $C_m$, which is logarithmic to all orders,
explaining why \eq{d1QQ_bT} has to have
the simple logarithmic mass dependence that we found for $\zb \to 1$.
We stress that starting at $\ord{\as^2}$, the soft and collinear-soft functions
\emph{do} develop a highly nontrivial dependence on $m b_T$,
which is thus inherited by the coefficients of $\cL_0(\zb)$ and $\delta(\zb)$
in $d_{1\,Q/Q}^{(2)}$ at two loops.
The required ingredients are readily available in the literature~\cite{Becher:2005pd, Pietrulewicz:2017gxc},%
\footnote{The secondary mass effects in $\sqrt{S}$ have been calculated in \refcite{Pietrulewicz:2017gxc}.
For the collinear-soft function one may use that for the exponential regulator of \refcite{Li:2016axz},
the collinear-soft function is simply given by an exponential
of the Collins-Soper kernel to all orders~\cite{Billis:2019vxg}.
This remains true in the presence
of mass effects in the CS kernel, which have likewise been calculated in \refcite{Pietrulewicz:2017gxc}.
One potential subtlety is that the results of \refcite{Pietrulewicz:2017gxc} for $\sqrt{S}$
were obtained using the $\eta$ regulator of \refscite{Chiu:2011qc, Chiu:2012ir} instead.
However, it is easy to see that the one-loop soft function
with a massive gluon (which underlies the dispersive calculation
of the secondary quark mass effects in \refcite{Pietrulewicz:2017gxc})
leads to identical results for both regulators,
meaning the final renormalized result
for the mass-dependent soft function in \refcite{Pietrulewicz:2017gxc}
is also valid for either regulator.
}
which can provide a strong cross check
on a future two-loop calculation of the partonic heavy-quark TMD FF,
or simplify it by allowing to restrict to finite $z < 1$ from the start.

\section{Application: EEC with heavy quarks in the back-to-back limit}
\label{sec:eec}

\subsection{Mass-dependent EEC jet functions}
\label{sec:eec_jet_functions}

In this section we use our results to produce explicit results
for the effect of heavy quarks
on the energy-energy correlator (EEC) observable~\cite{Basham:1978bw}
as measured in $e^+e^-$ collisions,
\begin{align} \label{eq:def_eec}
\mathrm{EEC}(\chi)
\equiv \frac{1}{\sigma_\mathrm{total}} \, \frac{\df \sigma} {d \cos\chi}
= \frac{1}{\sigma_\mathrm{total}} \sum_{a,b\, \in X} \int \! \df \sigma_{ee \to X}
\frac{E_a E_b}{Q^2} \delta \bigl[ \cos (\theta_{ab}) - \cos(\chi) \bigr]
\,,\end{align}
where the sum runs over all pairs of hadrons $a$ and $b$ in the event
with energies $E_a, E_b$ and $Q$ is the center-of-mass energy.
We are specifically interested in the so-called back-to-back limit $\chi \to \pi$,
where a parton-level factorization theorem has been proven~\cite{Moult:2018jzp}
and can be directly related to TMD factorization.
This precisely complements the study of \refcite{Craft:2022kdo},
where the NLO mass effects in the opposite (collinear) limit $\chi \to 0$ of the EEC were calculated,
and can serve as data for a future calculation of the heavy-quark EEC
for general angles.

To do so, it is convenient to rewrite the EEC as a differential distribution
in $z_\chi = \bigl[ 1 - \cos(\theta_{ab}) \bigr]/2 \to 1$,
i.e., the average pairwise angle of energy-weighted hadron pairs,
$\mathrm{EEC}(\chi) = (1/\sigma_\mathrm{total}) \, \df \sigma/\df z_\chi$.
We are interested in the modifications of the fixed-order EEC
from the presence of heavy quarks in this limit,
counting $(1 - z_\chi)Q^2 \sim 1/b_T^2 \sim m^2$,
where $b_T$ -- precisely as in TMD factorization --
is the typical transverse spacetime separation
of fields in the relevant correlators.
The $z_\chi$-differential distribution is given
by $z_\fh$-moments of the TMD FFs for identified hadrons~\cite{Moult:2018jzp},
\begin{align} \label{eq:eec_fact}
\frac{1}{\sigma_0} \frac{\df \sigma}{\df z_\chi}
&= \sum_{i,j} \cH_{ee\to ij}(Q^2, \mu) ~
Q^2 \!\! \int_0^\infty \! \df b_T \, b_T \, J_0\bigl( \sqrt{1-z_\chi} \, b_T Q \bigr)
\\
& \quad \times \sum_{\fh, \fh'}
\int \! \df z_\fh \, z_\fh^3 D_{1\,\fh/i}(z_\fh, b_T, \mu, Q^2)
\int \! \df z_{\fh'} \, z_{\fh'}^3 D_{1\,\fh'/j}(z_{\fh'}, b_T, \mu, Q^2)
\nn \\
&\equiv \sum_{i,j} \cH_{ee\to ij}(Q^2, \mu) ~
Q^2 \!\! \int_0^\infty \! \df b_T \, b_T \, J_0\bigl( \sqrt{1-z_\chi} \, b_T Q \bigr)
\nn \\ & \quad \times
J_i(b_T, m, \mu, Q/\nu) \,
J_{j}(b_T, m, \mu, Q/\nu) \,
S(b_T, m, \mu, \nu)
\nn \,,\end{align}
where the sum runs over all possible pairs of partons $i, j$,
with nonzero contributions for $i, j = q, \bar q$ a light quark pair
or $i, j = Q, \bar Q$ a heavy quark pair,
and it is understood that $\cH_{ee\to gg} = 0$.
Here our normalization is such that $\df \sigma/\df z_\chi = \sigma_0 \, \delta(1 - z_\chi) + \ord{\as}$
in the limit where the process is mediated only by virtual photons,
and the hard function $\cH_{ee\to ij}$ is given in \app{perturbative_ingredients}.
On the second equality we defined the EEC back-to-back jet function
$J_i(b_T, m, \mu, \nu)$ for a -- possibly massive -- parent parton $i$,
and we followed the usual convention in the EEC literature
to factor off a rapidity and UV-renormalized TMD soft function $S(b_T, m, \mu, \nu)$.
Both $J_i$ and $S$ acquire mass dependence for $1/b_T^2 \sim m^2$
from primary and secondary heavy-quark effects,
which we now classify and evaluate further.
Note that while the energy fractions entering \eq{def_eec}
in general differ from the lightcone momentum fractions $z_\fh$
that the TMD FFs depend on, in particular for massive particles,
the difference is power-suppressed in the back-to-back limit
where $k_T, m, m_\fh \ll z_\fh Q$.

To evaluate the sums over hadrons, we break each of them down
into two terms, a sum over light hadrons $\fh = h$
and one over heavy hadrons $\fh = H$.
For light-hadron final states, the standard manipulations
of the collinear FF using its momentum sum rule apply~\cite{Moult:2018jzp},
such that also in the presence of intermediate heavy quarks
and for all of $i = q, Q, g$,
\begin{align} \label{eq:eec_hadron_sum_light}
&\sum_{h} \int \! \df z_h \, z_h^3 \, D_{1\,h/i}(z_h, b_T, \mu, Q^2)
\nn \\
&= \sum_{k = q, \bar{q}, g} \int \! \df z \, z  \, \cJ_{k/i}(z, b_T, m, \mu, Q^2)
+ \ord{\lqcd^2 b_T^2}
\,,\end{align}
where $\cJ_{1\,k/i}$ is the mass-dependent matching coefficient
introduced in \eq{nonvalence_matching_light_hadron}.
Importantly, at the scale $\mu \sim \lqcd$ light partons
can only fragment to light hadrons
in the theory with $n_\ell$ light quark flavors,
which means that the collinear FF momentum sum rule
holds for the set of light hadrons $\{h\}$ on their own within that theory.

On the other hand, for heavy-hadron final states $H$
containing either $Q$ or $\bar{Q}$, we have
\begin{align} \label{eq:eec_hadron_sum_heavy}
&\sum_H \int \! \df z_H \, z_H^3 \, D_{1\,H/i}(z_H, b_T, \mu, Q^2)
\nn \\
&= \sum_{K = Q, \bar{Q}} \int \! \df z \, z^3 \, d_{1\,K/i}(z, b_T, \mu, Q^2)
+ \ord{\lqcd^2 b_T^2}
\,,\end{align}
which again holds for all of $i = q, Q, g$.
While the end result looks similar to \eq{eec_hadron_sum_light},
the moment of the perturbative coefficient arises
in a different (and much simpler) way,
namely because the fragmentation probabilities $\chi_H$
from \eq{tmd_ff_unpol_bhqet} sum to unity, while $d_{1\,Q/i}$ and $d_{1\,\bar{Q}/i}$
are independent of the state $H$.
We note in passing that bound states containing multiple heavy quarks
also in general contribute to the sum over hadrons in \eq{eec_fact}
and e.g.\ for a $Q\bar{Q}$ bound state can be evaluated
in terms of the TMD FF matching coefficients
onto NRQCD long-distance matrix elements calculated
in \refscite{Echevarria:2020qjk, Copeland:2023wbu, Echevarria:2023dme},
but this contribution to the jet function $J_i$ only starts at $\ord{\as^2}$.

The expressions above are particular ($N = 1$) Mellin moments of the partonic heavy-quark TMD FF
and the TMD FF matching coefficients,
which we in general define as
\begin{align} \label{eq:def_mellin_moments_bT}
d_{1\,Q/i}^{[N]}(b_T, \mu, \zeta)
&\equiv \int \! \df z \, z^{N + 2} d_{1\,Q/i}(z, b_T, \mu, \zeta)
\,, \nn \\
\cJ_{k/i}^{[N]}(b_T, m, \mu, \zeta)
&\equiv \int \! \df z \, z^{N}
\cJ_{k/i}(z, b_T, m, \mu, \zeta)
\,,\end{align}
where for $d_{1\,Q/i}$ we explicitly compensate
for the usual measure factor of $z^2$
when taking moments at fixed $b_T$ (or $k_T$).
In summary, we have
\begin{align}
&J_i(b_T, m, \mu, Q/\nu) \bigl[ S(b_T, m, \mu, \nu) \bigr]^{1/2}
\nn \\
&= \sum_{K = Q, \bar{Q}} d_{1\,K/i}^{[1]}(b_T, \mu, Q^2)
+ \sum_{k = q, \bar{q}, g} \cJ_{k/i}^{[1]}(b_T, m, \mu, Q^2)
\,,\end{align}
for the quark, massive quark, or gluon EEC back-to-back jet functions,
which generalizes the defining relation for the massless jet function in \refcite{Moult:2018jzp}.
We collect the Mellin moments of our new results
that enter the right-hand side in \app{mellin_moments}.
We point out that the organization of our final one-loop result for the heavy-quark
TMD FF in momentum space in terms of plus distributions
on rectangular domains in $(z, k_T)$, see \eq{d1qq_kT_ren_one_loop_result},
makes it particularly straightforward to take moments with respect to $z$.

\subsection{Jet function consistency in the small and large-mass limits}
\label{sec:eec_jet_functions_consistency}

The all-order consistency relation for the mass-dependent EEC jet functions
in the small-mass limit follows from \eq{consistency_relations_small_mass} and reads
\begin{align} \label{eq:eec_jet_func_small_mass}
J_i(b_T, m, \mu, Q/\nu) = J_i(b_T, \mu, Q/\nu) + \ord{m b_T}
\,,\end{align}
i.e., they simply reduce to their massless counterparts
because the partonic heavy-quark collinear FFs
and the FF flavor matching kernels preserve
the momentum sum rule.
(In other words, the measurement becomes fully inclusive
over the details of the IR also in this case, which now includes the mass.)

On the other hand, in the large-mass limit we can infer
from the discussion in \sec{theory_large_mass}
that the heavy-quark jet function to all orders satisfies
\begin{align} \label{eq:eec_jet_func_large_mass}
J_Q(b_T, m, \mu, Q/\nu) \bigl[ S(b_T, m, \mu, \nu) \bigr]^{1/2}
= C_m(m, \mu, Q^2) \, C_1\Bigl( b_T, \mu, \frac{Q}{m} \Bigr)
\,.\end{align}
Note that it is not power suppressed in this regime
because even at scales below the quark mass,
the heavy quark continues to radiate light modes and pick up recoil,
leading to a nontrivial angular dependence, and thus a nonzero EEC.
As for the TMD FFs, the mass effects
in the light-parton jet functions $J_q$ and $J_g$
for $(1-z_{\chi})Q^2 \ll m^2$ are of purely virtual origin
and encoded in collinear and soft mass matching coefficients
onto their counterparts in the $n_\ell$ (purely light) theory,
see \refscite{Pietrulewicz:2017gxc, Pietrulewicz:2023dxt}.

\subsection{Complete mass dependence of the EEC at \texorpdfstring{$\ord{\as}$}{O(as)}}
\label{sec:eec_complete_nlo}

The mass effects in the light-quark mediated cross section $i = q, \bar{q}$
only enter through closed heavy-quark loops or pair production starting at $\ord{\as^2}$,
such that at $\ord{\as}$ the total mass effect simply reads
\begin{align}
\frac{1}{\sigma_0} \frac{\df \sigma^{(1)}}{\df z_\chi}
&= \frac{1}{\sigma_0} \frac{\df \sigma^{(1)}_{ee \to q\bar{q}}}{\df z_\chi}
+ 2 \cH_{ee\to Q\bar{Q}}^{(1)}(Q^2, \mu) \, \delta(1-z_\chi)
\\ & \quad
+ 4 \cH_{ee\to Q\bar{Q}}^{(0)} \, \pi Q^2 \Bigl\{
   d_{1\,Q/Q}^{[1]\,(1)}\bigl(\sqrt{1-z_\chi} \, Q, \mu, Q^2 \bigr)
   + \cJ_{g/Q}^{[1]\,(1)}\bigl(\sqrt{1-z_\chi} \, Q, \mu, Q^2 \bigr)
\Bigr\}
\nn \,,\end{align}
where $\df \sigma_{ee \to q\bar{q}}$ refers to the light quark-mediated cross section,
which is known to $\ord{\as^3}$~\cite{Ebert:2020sfi},
and the expression for the hard function $\cH_{ee\to Q\bar{Q}}$
is given in \app{perturbative_ingredients}.
We also exploited that the convolution (or the inverse Fourier transform)
is trivial at fixed $\ord{\as}$,
and that the contributing partonic TMD FFs are all equal.
Using the results of \app{mellin_moments} for $N = 1$
and adding the contribution from the one-loop hard-function, we find
\begin{align}
\frac{1}{\sigma_0} \frac{\df \sigma^{(1)}}{\df z_\chi}
&= \frac{1}{\sigma_0} \frac{\df \sigma^{(1)}_{ee \to q\bar{q}}}{\df z_\chi}
+ 2\cH_{ee\to Q\bar{Q}}^{(0)} \, \frac{\as C_F}{4\pi} \biggl\{
8(2 \ln \rho-1) \, \cL_0(\zb_\chi)+ 8\Bigl(2\ln ^2 \rho +\ln \rho +\frac{2 \pi^2}{3}  - 2\Bigr)\,  \delta(\zb_\chi)
\nn \\ & \qquad
+ 4 \rho^2 \frac{
   \pi (-3 - 18 x + x^2) + \sqrt{x} (9 + x - 9x^2 - x^3) - \sqrt{x} (1 + 24x + 9x^2 + 2x^3)\ln x
}{
   \sqrt{x} (1+x)^4
}
\biggr\}
\nn \\ & \quad
+ \ord{\as^2} + \Ord{\zb_\chi^0} + \ORd{\frac{1}{\rho}}
\,,\end{align}
where we used the shorthands $\rho \equiv Q/m$
for the boost of the heavy quarks,
$\zb_\chi \equiv 1 - z_\chi$, and $x \equiv \rho^2 \zb_\chi = k_T^2/m^2$.
The result is valid up to power corrections in $1/Q$, i.e., $\ord{\zb_\chi^0}$
or $\ord{1/\rho}$, but captures the exact dependence on $\rho^2 \zb_\chi \sim 1$.

\subsection{Extension to nonperturbative \texorpdfstring{$1 - z_{\chi,H} \sim \lqcd^2/Q^2$}{1-zchiH sim LQCD2/Q2}}
\label{sec:eec_nonperturbative}

To close this section, we note that an intriguing modification
to the observable can be made
where the sum in the \emph{observable definition} in \eq{def_eec}
is restricted to heavy hadrons $a, b \in \{H\}$ containing valence heavy quarks.
For this modified observable, which we may call $\mathrm{EEC}_H$
with an associated differential variable $z_{\chi,H}$,
it is in fact straightforward to extend the factorization
all the way down into the nonperturbative region $1 - z_{\chi,H} \sim \lqcd^2/Q^2$:
\begin{align} \label{eq:eec_h_fact_nonpert}
\frac{1}{\sigma_0} \frac{\df \sigma}{\df z_{\chi,H}}
&=\cH_{ee\to Q\bar{Q}}(Q^2, \mu) \, C_m^2(m, \mu, Q^2)
\nn \\ & \quad \times
Q^2 \!\! \int_0^\infty \! \df b_T \, b_T \, J_0\bigl( \sqrt{1-z_{\chi,H}} \, b_T Q \bigr)
\Bigl[ \chi_1\Bigl(b_T, \mu, \frac{Q}{m}\Bigr) \Bigr]^2
\,,\end{align}
where $\chi_1$ is the renormalized Wilson-line correlator
defined in \eq{def_chi1_ren}.
The simplicity of \eq{eec_h_fact_nonpert},
which is valid for $\lqcd \lesssim \sqrt{1 - z_{\chi,H}} \, Q \ll m \ll Q$
and coincides with \eq{eec_jet_func_large_mass} for $\lqcd \ll \sqrt{1 - z_{\chi,H}} \, Q$,
comes about exactly as for the individual $\chi_1$ in \sec{bhqet_tmd_fragmentation_factors}:
Summing over heavy hadrons $H$ allows one
to complete the sum over states in the EFT
and reduces the final result to (a square of) vacuum correlators
of Wilson lines.
We note that in \refcite{vonKuk:2023jfd}, we already mentioned the factorization
in \eq{eec_h_fact_nonpert} as a factorization theorem for the total heavy-hadron
TMD cross section in $e^+ e^-$ collisions,
but that -- in hindsight -- of course coincides
with the EEC in the back-to-back limit.

\section{Comparison to \texorpdfstring{\href{https://arxiv.org/abs/2310.19207}{2310.19207}}{2310.19207} by Dai, Kim, and Leibovich}
\label{sec:comparison_to_2310_19207}

Recently, an independent NLO calculation of the heavy-quark TMD FFs
$d_{1\,Q/Q}$ and $d_{1\,Q/g}$ was performed in \refcite{Dai:2023rvd}.
Before commenting on the differences between the two calculational approaches,
we first want to stress that after accounting for a measure factor of $1/z^2$
and the renormalized TMD soft function in \eq{tmd_s_cmu_one_loop},
we find complete agreement of our \eqs{d1QQ_bT}{nonvalence_bT_one_loop}
with the final $b_T$-space results
given in their eqs.~(2.36) and (4.26).
In order to gauge the strength of this cross check,
it is relevant to note that while the calculation of the real-emission
diagrams in \refcite{Dai:2023rvd},
like ours in \sec{tmd_ff_reals_and_expansion},
was initially performed in momentum space and only Fourier-transformed
after renormalization,
\refcite{Dai:2023rvd} used a completely orthogonal method
to isolate the singularities in $d^{(1)}_{1\,Q/Q}$,
i.e., to address the key source of subtleties.
Specifically, in \refcite{Dai:2023rvd} the characteristic denominator
$1/X \equiv 1/(k_T^2 + m^2 \zb^2/z^2)$ appearing in the real-emission diagrams
was directly expanded into one-dimensional plus distributions
of $X$ and in addition was multiplied by one-dimensional distributions of $1-z$,
which effectively amounts to performing a change of variables to $(z, X)$.
This procedure is completely orthogonal to our approach,
namely, we expand in terms of one-dimensional plus distributions
along the boundaries $\zb = 0$ and $k_T = 0$
and a two-dimensional plus distribution
whose defining domain \emph{aligns} with the main variables $(z, k_T)$.
The two approaches in particular differ (and have to differ)
for the boundary contribution proportional to $\delta(1-z)$.
Indeed, the process in \refcite{Dai:2023rvd}
generates highly nontrivial twofold integrals $F(m^2/\Lambda^2)$
and $G(m^2/\Lambda^2)$ multiplying $\delta(1-z)$,
where $\Lambda$ is the boundary condition of the $X$ plus distributions,
with no closed form for $F$ and $G$ away from asymptotic limits in $m$.

In view of these substantial differences, it is highly nontrivial
that the final results in $b_T$ space are indeed in full agreement,
in particular regarding the latter's $\delta(1-z)$ boundary condition.
In practical terms, we believe that our final momentum-space
result in \eq{d1qq_kT_ren_one_loop_result}
compared to the one in eq.~(2.30) of \refcite{Dai:2023rvd} nevertheless has
significant advantages.
A first, practical advantage is that \eq{d1qq_kT_ren_one_loop_result}
is expressed in closed form
in terms of elementary functions.
A perhaps more important second advantage is
that aligning the domains of the plus distributions
with the main variables of interest
makes it very straightforward to take integral transforms
with respect to either $k_T$ or $z$,
see the discussion above \eq{plus_distros_2d_how_to_take_x_cumulant}
and \eq{mellin_moment_2d_plus}.

\Refcite{Dai:2023rvd} also considered how the heavy-quark TMD FF
factorizes in the bHQET limit $k_T \ll m$.
At variance with our assumption of $1 - z_H \sim 1$
in \refcite{vonKuk:2023jfd} and in \eq{how_we_count_zH},
which amounts to considering wide bins or moments of $z_H$,
\refcite{Dai:2023rvd} counts $1 - z_H \sim k_T/m \ll 1$ in the bHQET regime,
i.e., staying differential in $z_H$ but restricting to the peak region,
which leads to a TMD bHQET shape function that depends on $z_H$,
cf.\ the comment below eq.~(2.6) in our \refcite{vonKuk:2023jfd}.
Again, before commenting on the differences between our approaches further,
we want to stress that for the power counting of $1-z_H \sim k_T/m$ assumed
by the authors, their factorization of the hadron-level TMD FF
in terms of $C_m$ and their TMD shape function
(which may further be matched onto a collinear shape function,
depending on the hierarchy between $\lqcd$ and $k_T$)
is entirely correct.

One difference between the two power-counting assumptions on $1-z_H$
regards the kind of cross check the respective bHQET calculation provides
on the large-mass behavior of the TMD FF.
The approach of \refcite{Dai:2023rvd} checks the differential $z$ dependence of $d_{1\,Q/Q}^{(1)}$,
but in intermediate steps an identical plus distribution
expansion with respect to a very similar, complicated denominator $1/X' = 1/(k_T^2 + \zb^2 m^2)$ is used,
which is subject to the same subtleties as the full calculation.
In this sense, our calculation provides a strong complementary check
of the large-mass behavior since it directly probes the total integral,
i.e., a defining property of the distributional result.%
\footnote{\Refcite{Dai:2023rvd} also performed a small-mass consistency check,
using the one-loop TMD PDF matching coefficients
and a reciprocity relation. This provides a sufficient check at one loop,
but reciprocity relations between TMD PDF and FF matching coefficients
are known to be subtle at higher orders~\cite{Ebert:2020qef, Luo:2020epw}.
Here we stress that the correct all-order relation for the small-mass limit
is the one we originally gave in \refcite{vonKuk:2023jfd},
i.e., it has to involve the genuine TMD \emph{FF} matching coefficients
onto twist-2 collinear FFs because the UV matching coefficients
are independent of the details of the IR,
that is, whether one matches onto a theory at $\mu \sim \lqcd$ or $\mu \sim m$.}

A second and final difference between the power counting assumptions
is phenomenological and regards the kinds of differential distributions
one may describe in either approach.
Our stance here is that while -- again -- the assumption of \refcite{Dai:2023rvd}
is sufficient and entirely correct to describe specific regions
of the multi-differential $(z_H, k_T)$ distribution,
it is not the unique power-counting assumption
on $1 - z_H$, $k_T$, and $m$
that would yield a leading-power factorization theorem involving bHQET modes,
in particular not when different possible scalings of $\lqcd$ are also considered.
As one example, we point to our analysis in \sec{joint_res},
where we continued to count $m \sim k_T$, but still took $1 - z_H$ small,
and likewise obtained a leading-power factorization theorem,
in that case also involving a purely collinear bHQET shape function,
but nontrivial $k_T$-dependent correlators of lightlike Wilson lines at the scale $\mu \sim k_T$.
In light of this, we believe that capturing the full $(z_H, k_T)$ dependence
will require dedicated future studies combining all of these results,
while for the time being and within the scope of exploratory theory studies
(or, necessarily less differential, initial experimental measurements)
the restriction to wide bins or low-moment integrals over $z_H$ is reasonable.

\section{Conclusions}
\label{sec:conclusions}

In this paper, we calculated all heavy-quark transverse-momentum dependent
fragmentation functions (TMD FFs) to next-to-leading order (NLO)
in the strong coupling.
We provided explicit results
in transverse momentum ($k_T$) space
-- both for the differential and the cumulative distribution --
and in the conjugate position ($b_T$) space,
retaining the exact dependence on $k_T/m$ and $b_T m$, respectively,
with $m$ the mass of the heavy quark.
Our calculations provide the last missing, key ingredient
for a complete next-to-next-to-leading logarithmic (NNLL)
description of the transverse momentum distributions
of observed heavy hadrons, capturing all quark mass effects.
Our results also enable the extension of fixed-order subtraction methods
to quasi-collinear limits involving final-state heavy quarks.

We further studied the corresponding TMD fragmentation factors
and the TMD light-spin density matrix
that arise in boosted Heavy-Quark Effective Theory (bHQET)
when the transverse momentum is small compared to the mass.
We demonstrated explicitly that they satisfy a two-dimensional set
of renormalization group equations (RGEs)
with respect to the standard virtuality scale $\mu$
and a dimensionless analog $\rho = \sqrt{\zeta}/m$ of the usual Collins-Soper scale $\zeta$.
Here $\rho$ has an immediate physical interpretation as the boost
of the heavy hadron in the lab frame.
We then proceeded to also evaluate the unpolarized TMD fragmentation factor
at NLO, allowing us to cross check our results for the heavy-quark TMD FF
in the large-mass and -- using known results from the literature --
the small-mass limit, finding full agreement.
We further showed how to extend the theory of joint threshold and $k_T$ resummation
to fully account for quark mass effects of $\ord{m^n b_T^n}$,
and used these novel factorization theorems for the heavy-quark TMD FF
to obtain a third independent cross check on our NLO result in an asymptotic limit.
We also compared our final results for the exact heavy-quark TMD FF
in $b_T$ space to a recent independent calculation
by another group. We found full agreement despite the highly orthogonal ways
in which singularities were isolated in the two intermediate momentum-space calculations,
which provides a strong independent cross check of this key universal ingredient
for heavy-quark TMD predictions.
Finally, and as a first application,
we leveraged our results to obtain the exact quark-mass dependence
of the energy-energy correlator (EEC) in $e^+e^-$ collisions in the back-to-back limit,
where the associated jet functions are given by Mellin moments of our results.
As a side note, we pointed out that if the EEC
is restricted to heavy hadrons ($\mathrm{EEC}_H$),
substantial simplifications occur in the back-to-back limit
such that the $\mathrm{EEC}_H$ at values below the quark mass
is given by a simple product of Wilson-line vacuum expectation values,
even extending down to nonperturbative values of the observable.

Our results strengthen the theoretical underpinnings of heavy-quark TMD FFs
and pave the way towards precision studies of their nonperturbative dynamics,
as well as for improving the description of heavy flavor differential distributions
in general-purpose Monte-Carlo programs.
We look forward to future applications of our results
in precision predictions for transverse momentum distributions
of heavy quarks and hadrons,
and to comparing them to future experimental measurements
to shed light on the three-dimensional dynamics of heavy-quark hadronization.

\acknowledgments

We thank Iain Stewart, Kyle Lee, and Maximilian Stahlhofen for helpful discussions.
We thank the Erwin Schr\"{o}dinger Institute for hospitality
while parts of this work were performed.
This work was supported in part by the Office of Nuclear Physics of the U.S.\
Department of Energy under Contract No.\ DE-SC0011090.
Z.S. was also supported by a fellowship from the MIT Department of Physics.
R.v.K. was supported by the European Research Council
(ERC) under the European Union’s Horizon 2020 research and innovation programme
(Grant agreement No. 101002090 COLORFREE).
J.M. was supported by the D-ITP consortium, a program of NWO that is funded by the Dutch Ministry of Education, Culture and Science (OCW).

\appendix

\section{Notation and conventions}
\label{app:conventions}

\subsection{Lightcone coordinates}
\label{app:lc_coordinates}

We decompose four-vectors $p^\mu$
in terms of lightlike vectors $n^\mu$, $\bn^\mu$ with $n^2 = \nbar^2 = 0$
and {\boldmath $n \cdot \bn = 2$},
\begin{align}
p^\mu
= \nbar \cdot p \, \frac{n^\mu}{2} + n \cdot p \, \frac{\nbar^\mu}{2} + p_\perp^\mu
\equiv p^- \frac{n^\mu}{2} + p^+ \frac{\nbar^\mu}{2} + p_\perp^\mu
\equiv (p^-, p^+, p_\perp)
\,,\end{align}
such that e.g.\ $p^2 = p^- p^+ + p_\perp^2$.
We take collinear momenta to
have large components $p^- \gg p_\perp \gg p^+$.
We always take transverse vectors
with subscript $\perp$ to be Minkowskian, $p_\perp^2 \equiv p_\perp \cdot p_\perp \leq 0$,
and denote their magnitude by $p_T = \sqrt{- p_\perp^2}$
such that $p_T^2 \geq 0$ is the square of a real number.

\subsection{Plus distributions}
\label{app:plus_dist}

One-dimensional plus distributions are defined through their action
on test functions $g(x)$,
\begin{align} \label{eq:def_plus}
\int \! \df x \bigl[f(x) \bigr]_+ \, g(x)
\equiv \int \! \df x \, f(x) \bigl[ g(x) - g(0) \, \Theta(1 - x) \bigr]
\,,\end{align}
where $f(x)$ diverges at most as $x^{-1-\alpha}$ with $\alpha < 1$ for $x \to 1$
and $g(x)$ is assumed to be differentiable as $x \to 0$.
Integrals of $[f(x)]_+$ against the constant test function
over the domain $[0,1]$ vanish by construction.
The generalization of \eq{def_plus} to two dimensions
is given in \eq{def_plus_plus} in the main text.
For logarithmic plus distributions with homogeneous power counting $\sim 1/x$
and their analogs for dimensionful variables,
we further define the shorthands
\begin{align} \label{eq:def_cLn}
\cL_n(x) \equiv \Bigl[ \frac{\Theta(x)\,\ln^n x}{x} \Bigr]_+
\,, \quad
\cL_n(k, \mu) \equiv \frac{1}{\mu} \cL_n\Bigl( \frac{k}{\mu} \Bigr)
\,, \quad
\cL_n(k_T^2, \mu^2) \equiv \frac{1}{\mu^2} \cL_n\Bigl( \frac{k_T^2}{\mu^2} \Bigr)
\,.\end{align}
The definition in \eq{def_plus}
enables expansions in a regulator $\alpha \sim \eps, \eta$
as follows,
\begin{alignat}{3}
\label{eq:expanding_in_plus_distros}
\frac{\Theta(x)}{x^{1+\alpha}}
&= -\frac{ \delta (x)}{\alpha}
+ \Bigl[ \frac{\Theta(x)}{x^{1+\alpha}} \Bigr]_+
&&= -\frac{ \delta (x)}{\alpha} + \cL_0(x) - \alpha \cL_1(x) + \ord{\alpha^{2}}
\,, \\
\label{eq:expanding_in_plus_distros_with_log}
\frac{\Theta(x) \, \ln x}{x^{1+\alpha}}
&= -\frac{ \delta (x)}{\alpha^2}
+ \Bigl[ \frac{\Theta(x) \, \ln x}{x^{1+\alpha}} \Bigr]_+
&&= -\frac{ \delta (x)}{\alpha^2} + \cL_1(x) + \ord{\alpha}
\,.\end{alignat}
To relate plus distributions $\cL_n(k, \mu)$
with different boundary conditions $\mu_{1,2}$,
it is useful to shift one of the boundary conditions
by making use of identities like
\begin{align} \label{eq:shift_bc}
\cL_0(k, \mu_2)
&= \cL_0(k, \mu_1)
+ \delta(k) \, \ln \frac{\mu_1}{\mu_2}
\,, \nn \\
\cL_1(k, \mu_2)
&= \cL_1(k, \mu_1)
+ \cL_0(k, \mu_1) \, \ln \frac{\mu_1}{\mu_2}
+ \delta(k) \, \frac{1}{2} \ln^2 \frac{\mu_1}{\mu_2}
\,.\end{align}

\section{Perturbative ingredients}
\label{app:perturbative_ingredients}

In this appendix we collect various perturbative ingredients
from elsewhere in the literature in our notation,
as used for cross-checks of our results in the main text.
The following expressions are all accurate to one-loop order.

\paragraph{Hard function.} The hard function describing the pair production of quarks
through a virtual photon or $Z$ boson in $e^+e^-$ collisions
is given by
\begin{align} \label{eq:hard_ee_hadrons_tree}
\cH_{ee \to ij}(Q^2, \mu)
&= \delta_{i \bar{j}} \,
\Bigl\{
e_i^2
- 2 v_e v_i e_i \Re \bigl[ P_Z(Q^2) \bigr]
+ (v_e^2 + a_e^2)(v_i^2 + a_i^2) \Abs{P_Z(Q^2)}^2
\Bigr\} \Abs{C_q}^2
\,, \nn \\
C_q(Q^2, \mu) &= 1 + \frac{\as C_F}{4\pi} \Bigl[ -\ln^2 \frac{-Q^2 -\img 0}{\mu^2} + 3 \ln \frac{-Q^2 -\img 0}{\mu^2} - 8 + \frac{\pi^2}{6} \Bigr]
\,.\end{align}
Here we have kept the contribution from $Z$ boson exchange
and $Z$-photon interference, as relevant for measurements on the $Z$ pole,
where $P_Z(Q^2) = Q^2/(Q^2 - m_Z^2 + \img \Gamma_Z m_Z)$ is the reduced $Z$ propagator
and $e_f$ ($v_f$, $a_f$) are the electromagnetic charge (vector, axial coupling to the $Z$)
of a fermion $f$.

\paragraph{Large-mass and joint-resummation limits.}

The combined collinear and soft mass matching coefficient $C_m$
see also \refcite{vonKuk:2023jfd} for our notation,
is given by \cite{Hoang:2015vua}
\begin{align}\label{eq:wilson_coeff_Cm}
C_m(m, \mu, \zeta) = 1 +\frac{\as C_F}{4 \pi} \biggl(
   4 \ln^2 \frac{\mu}{m} + 2\ln \frac{\mu}{m}+ 4 + \frac{\pi^2}{6}
\biggr) + \Ord{\as^2}
\,.\end{align}
At this order, it is also equal to the matching coefficient
$H_{m,n}^{Q\to h_v}(m, \mu, \nu/\sqrt{\zeta})$
(defined in \refcite{Hoang:2015vua}, without the superscript)
for matching massive collinear quark modes with label momentum $\sqrt{\zeta}$
onto bHQET on their own,
whereas the additional contribution $\sqrt{H_{m,s}}(m, \mu, \nu)$
from matching secondary soft heavy quarks onto light soft modes
becomes nontrivial at $\ord{\as^2}$.
Two-loop expressions for both are given in \refcite{Hoang:2015vua},
where they were extracted from the NNLO heavy-quark form factor~\cite{Bernreuther:2004ih, Gluza:2009yy}.
At that order, there is also a nonzero $\zeta$ dependence
that arises from the rapidity renormalization of $H_{m,n}^{Q\to h_v}$ and $H_{m,s}$.
The two-loop expression for secondary heavy-quark effects
on the light-to-light collinear matching coefficient $H_{m,n}^{q \to q}$
is given in \refcite{Pietrulewicz:2017gxc} as $H_c$,
where it was extracted from the results of \refscite{Pietrulewicz:2014qza, Hoang:2015vua}.
The perturbative HQET shape function is given by \cite{Neubert:2007je, Fickinger:2016rfd}
\begin{align} \label{eq:hatf_one_loop}
\hat{F}\Bigl( \frac{k^-}{v^-}, \mu \Bigr)
= \delta\Bigl( \frac{k^-}{v^-} \Bigr)
+ \frac{\as C_F}{4\pi} \Bigl[
   -8\cL_1\Bigl( \frac{k^-}{v^-}, \mu \Bigr)
   -4\cL_0\Bigl( \frac{k^-}{v^-}, \mu \Bigr)
   - \frac{\pi^2}{6} \delta\Bigl( \frac{k^-}{v^-} \Bigr)
\Bigr]
\end{align}
The renormalized TMD soft function for both the $\eta$ regulator of \refcite{Chiu:2011qc, Chiu:2012ir}
and the exponential regulator of \refcite{Li:2016axz} reads
\begin{align} \label{eq:tmd_s_cmu_one_loop}
   \sqrt{S}(b_T, m, \mu, \nu)
   = 1 + \frac{\as C_F}{4\pi} \Bigl[ -L_b^2 + 4 L_b \ln \frac{\mu}{\nu} - \frac{\pi^2}{6} \Bigr]
\,.\end{align}
The joint-resummation collinear-soft function is given by~\cite{Lustermans:2016nvk}
\begin{align} \label{eq:joint_res_csoft_one_loop}
   \cS(\ell^-, b_T, m, \mu, \nu)
   = \delta(\ell^-) + \frac{\as C_F}{4\pi} \Bigl[ -8 L_b \, \cL_0(\ell^-, \nu) \Bigr]
\,.\end{align}

\paragraph{Small-mass limit.}

The massless TMD FF matching coefficients
are given by \cite{Echevarria:2014rua, Collins:1350496, Ebert:2020qef}
\begin{align} \label{eq:matching_coeff_light_quark_limit}
\cJ_{q/q}(z, b_T, \mu, \zeta)
&= \delta (1-z) + \frac{\as C_F}{4 \pi } \biggl[
   \frac{4  (1 + z^2) \ln(z)}{ 1-z}-{4 L_b \cL_0(1-z)}
\nn \\ & \quad
   +\delta (1-z) \Bigl(
      -{L_b^2 } - 2L_b \ln \frac{\zeta}{\mu^2} -\frac{\pi ^2 }{6 }
   \Bigr) +{2L_b  (z+1)} + 2(1-z)
\biggr]\,,\nn \\
\cJ_{g/q}(z, b_T, \mu, \zeta)
&= \frac{\as C_F}{4 \pi } \biggl[
   \frac{4 (z^2-2 z+2) \ln z}{ z}-\frac{2L_b
	(z^2-2 z+2)}{ z}+{2 z}
\biggr]\,,\nn \\
\cJ_{g/g}(z, b_T, \mu, \zeta)
&= \delta (1-z)+\frac{\as C_A}{4 \pi} \biggl[
   \delta (1-z) \Bigl(-{L_b^2 }- 2L_b \ln \frac{\zeta}{\mu^2} -\frac{\pi ^2}{6}\Bigr)-{4 L_b \cL_0(1-z)}
\nn \\ & \quad
+ \frac{8  (z^2-z+1)^2 \ln z}{ (1-z) z}
+ \frac{4L_b (z^3-z^2+2 z-1)}{ z}
\biggr]\nn \\
\cJ_{q/g}(z, b_T, \mu, \zeta)
&= \frac{\as T_F}{4 \pi} \bigl[
   2(4 z^2-4 z+2) \ln z - 2L_b (2 z^2-2 z+1)+4 (1-z) z
\bigr]
\,.\end{align}
The collinear heavy-quark FFs $d_{Q/Q}(z,\mu)$ and $d_{Q/g}(z,\mu)$
are given by \cite{Mele:1990yq}
\begin{align} \label{eq:collinear_hq_FF}
d_{Q/Q}(z,\mu) &= \delta(1-z)+ \frac{\as}{4 \pi} C_F \Bigl\{
\ln \frac{\mu^2}{m^2} \bigl[ 4 \cL_0(1-z)+3 \delta(1-z)-2 (z+1) \bigr]
\nn \\ & \quad
-4 \cL_0(1-z)-8 \cL_1(1-z)+4 \delta(1-z)+2 (z+1) \bigl[ 2 \ln (1-z)+1 \bigr]
\Bigr\}\,,\nn\\
d_{Q/g}(z,\mu)=& \frac{\as}{4 \pi} T_F \bigl[ 2z^2+ 2(1-z)^2 \bigr] \ln \frac{\mu^2}{m^2}\,,
\,.\end{align}
The decoupling kernels relating collinear FFs in theories
with $n_f=n_\ell +1$ and $n_\ell$ active flavors are given by~\cite{Cacciari:2005ry}
\begin{align}\label{eq:decoupling_kernel}
\cM_{g/g}(z,\mu) &=\delta(1-z) + \frac{\as T_F}{4 \pi} \, \delta(1-z) \Bigl( - \frac{4}{3} \Bigr) \ln \frac{\mu^2}{m^2} \nn\\
\cM_{g/Q}(z,\mu) &= \frac{\as}{4 \pi} C_F \frac{1+(1-z)^2}{z} \Bigl(
   2\ln \frac{\mu^2}{m^2}-2 -4 \ln z
\Bigr)
\,.\end{align}
In both \eqs{collinear_hq_FF}{decoupling_kernel} we have rewritten
the original results in terms of the minimal distributional basis
of $\delta(1-z)$ and $\cL_n(1-z)$ using \eq{s_r_decomposition}.

\section{Results for Mellin moments}
\label{app:mellin_moments}

We first consider the Mellin moments of the momentum-space
TMD FFs and matching coefficients, which are defined
in analogy to \eq{def_mellin_moments_bT} as
\begin{align} \label{eq:def_mellin_moments_kT}
d_{1\,Q/i}^{[N]}(k_T, \mu, \zeta)
&\equiv \int \! \df z \, z^{N + 2} d_{1\,Q/i}(z, k_T, \mu, \zeta)
\,, \nn \\
\cJ_{k/i}^{[N]}(k_T, m, \mu, \zeta)
&\equiv \int \! \df z \, z^{N}
\cJ_{k/i}(z, k_T, m, \mu, \zeta)
\,.\end{align}
For $d_{1\,Q/Q}$, the relevant two-dimensional plus distribution
simply turns into
\begin{align} \label{eq:mellin_moment_2d_plus}
\int_0^1 \! \df z \, z^N \,\bigl[ f(x, z) \bigr]_{+,+}
= \Bigl[ \int_0^1 \! \df z \, \bigl( z^N - 1 \bigr) f(x, z) \Bigr]_+
\,,\end{align}
which like the moments of the one-dimensional plus distribution in $z$
can readily be evaluated for general $N$.
The remaining one-dimensional plus brackets in \eq{mellin_moment_2d_plus}
act on the $x = k_T^2/m^2$ dependence
and can further be decomposed in terms of singular $1/x$ terms
and regular terms of $\ord{x^0}$ in analogy to \eq{s_r_decomposition},
which is compensated by an additional $\delta(x)$ boundary condition.
As an explicit result, we find for $N = 1$ at one loop as relevant for \sec{eec}:
\begin{align}
d_{1\,Q/Q}^{[1]\,(1)}(k_T, \mu, \zeta)
&= \frac{\as C_F}{4\pi}
\frac{1}{\pi} \biggl\{
      2 \ln \frac{\zeta}{\mu^2} \cL_0(k_T^2, \mu^2)
      + 2 \ln \frac{\mu^2}{m^2} \cL_0(k_T^2, m^2)
      - \ln^2 \frac{\mu^2}{m^2} \delta(k_T^2) + 3 \ln \frac{\mu^2}{m^2} \delta(k_T^2)
\nn \\ & \qquad
+ \frac{1}{m^2} \frac{
   3\pi (-3 - 18x + x^2)
   - \sqrt{x} (-25 + 9x + 45 x^2 + 11 x^3)
}{
   \sqrt{x} (1 + x)^4
}
\nn \\ & \qquad
+ \frac{1}{m^2} \frac{
   - 3 \sqrt{x} (1 + 24 x + 9x^2 + 2x^3) \ln x
}{
   \sqrt{x} (1 + x)^4
}
- 2\cL_0(k_T^2, m^2)
+ \delta(k_T^2) \Bigl( 4 + \frac{\pi^2}{6} \Bigr)
\biggr\}
\,, \nn \\
d_{1\,Q/g}^{[1]\,(1)}(k_T, \mu, \zeta)
&= \frac{\as T_F}{4\pi}
\frac{1}{\pi m^2} \frac{
   2x(6 + x) - 30 \sqrt{x} \arctan \sqrt{x} - 3(-6 + x) \ln(1 + x)
}{
   3x^3
}
\,, \nn \\
\cJ_{1\,g/Q}^{[1]\,(1)}(k_T, m, \mu, \zeta)
&= \frac{\as C_F}{4\pi}
\frac{1}{\pi m^2} \frac{
   2(1 + 4x)
}{
   3(1 + x)^2
}
\,.\end{align}

\begin{figure*}
\centering
\includegraphics[width=0.48\textwidth]{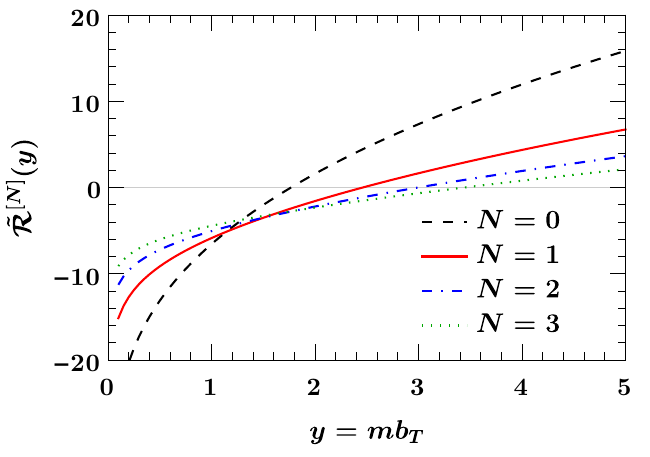}
\hfill
\includegraphics[width=0.48\textwidth]{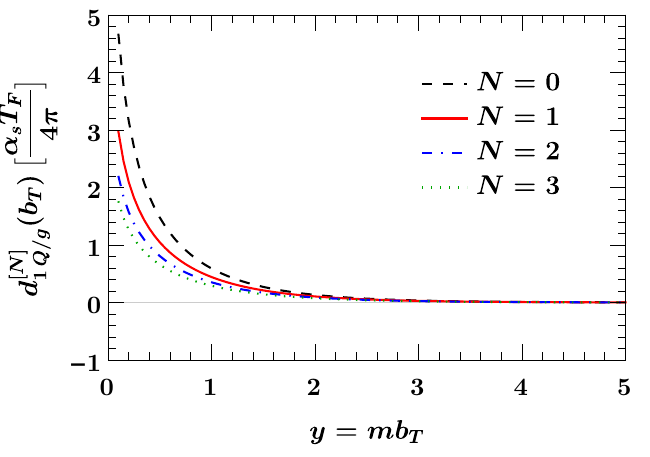}
\caption{
Left: Numerical Mellin moments $\cR^{[N]}(y)$
of the finite remainder term in $d_{1\,Q/Q}$.
Right: Numerical Mellin moments $d_{1\,Q/g}^{[N]\,(1)}$
of the gluon-to-massive-quark TMD FF.
}
\label{fig:numerical_mellin_moments}
\end{figure*}

Turning to the Mellin moments of the $b_T$-space TMD FFs in \eq{def_mellin_moments_bT},
we may use standard Mellin transforms of plus distributions
to transform the leading $z \to 1$ terms in $d_{1\,Q/Q}$ in \eq{d1QQ_bT},
\begin{align}
&d_{1\,Q/Q}^{[N]\,(1)}(b_T, \mu, \zeta)
\nn \\[0.2em]
&= \frac{\as C_F}{4\pi}
\bigg\{
- 2 L_b \ln\frac{\zeta}{m^2}
+ 4 \ln^{2}\frac{\mu}{m}
+ 6 \ln \frac{\mu}{m}
- L_y^2
+ 4 - \frac{\pi^{2} }{6}
+ 4 (1 + L_y ) \bigl[ \psi(1 + N) + \gamma_E \bigr]
\nn \\[0.2em]
& \qquad \qquad \quad
- 8 \Bigl[ -\tfrac{1}{2} \psi^{(1)}(1 + N) + \tfrac{1}{2} \bigl[ \psi(1 + N) + \gamma_E \bigr]^2 + \frac{\pi^2}{12} \Bigr]
+ \tcR^{[N]}(b_T m)
\bigg\}
\,,\end{align}
where $\psi(x)$ ($\psi^{(1)}(x)$) is (the first derivative of) the Digamma function.
We were not able to find a closed-form expression
for the Mellin transform
$\tcR^{[N]}(y) \equiv \int_0^1 \! \df z \, z^N \, \tcR(z, y)$
of the finite remainder term,
but implementing the $z$ integral of \eq{def_tcR} directly
leads to stable numerical results
(shown in the left panel of \fig{numerical_mellin_moments} for reference)
that could readily be interpolated.
For $\cJ_{g/Q}$, the Mellin transform for general $N$ is trivial
since $z$ does not appear in the arguments of the Bessel $K$ functions,
\begin{align}
\cJ_{g/Q}^{[N]\,(1)}(b_T, m) = \frac{\as C_F}{4\pi} \, 4\Bigl[
   \frac{N^{2} + 3N + 4}{N (N+1) (N+2)} K_{0}( b_{T}m)
   - \frac{b_{T} m}{2 N (N + 1)} K_{1}( b_{T}m)
   \Bigr]
\,.\end{align}
For $d_{1\,Q/g}^{[N]\,(1)}$, a closed-form expression
in terms of $_1 F_2$ hypergeometric and Meijer $G$ functions
may be obtained for general $N$ from its definition,
but is no more instructive than numerical $z$ moments of \eq{nonvalence_bT_one_loop}.
We show the latter in the right panel of \fig{numerical_mellin_moments} for reference.

\newpage
\addcontentsline{toc}{section}{References}
\bibliographystyle{jhep}
\bibliography{../paper/refs}

\end{document}